# Thermally enhanced magnon-phonon resonance in van der Waals antiferromagnet $FePS_3$

Kai-Xuan Zhang[1,4†*], Beom Hyun Kim[1†], Seungbok Lee[1], Hao Li[2], Eugene Park[1$], Heejun Yang[1,3], Feng Jin[2], Qingming Zhang[2*], and Je-Geun Park[1,4*]

[1]*Department of Physics and Astronomy, Seoul National University, Seoul 08826, South Korea*
[2]*Beijing National Laboratory for Condensed Matter Physics, Institute of Physics, Chinese Academy of Sciences, Beijing 100190, China*
[3]*RIKEN Center for Emergent Matter Science (CEMS), Wako, Saitama 351-0198, Japan*
[4]*KOMQUEST Center, Seoul National University, Seoul 08826, South Korea*
[$]Present address: Department of Physics, Massachusetts Institute of Technology, Cambridge, MA 02139, USA

[†] These authors contributed equally to this work.
[*]Corresponding authors: Kai-Xuan Zhang (email: kxzhang.research@gmail.com), Qingming Zhang (email: qmzhang@iphy.ac.cn), and Je-Geun Park (email: jgpark10@snu.ac.kr)

**Abstract**
Strong coupling between collective excitations provides a route to controlling quantum states of matter, yet its evolution under thermal fluctuations remains poorly understood. In the van der Waals antiferromagnet $FePS_3$, strong magnon-phonon coupling is established at low temperatures. At the same time, recent terahertz experiments revealed the counterintuitive stabilization of a long-lived metastable magnetic state with an exceedingly long lifetime of 5 msec at the antiferromagnetic transition, not at the base temperature. This was attributed to a strongly coupled spin-lattice scenario, challenging the conventional wisdom that increased magnetic and thermal fluctuations near $T_N$ should be detrimental to the coherent control of a metastable phase. Thus it requires an independent examination for correct understanding. Using high-field Raman spectroscopy, we systematically investigate the temperature evolution of magnon-phonon coupling in $FePS_3$. We observe a pronounced avoided crossing between magnon and phonon modes across a wide range of temperatures and find that the coupling strength remains remarkably robust with increasing temperatures. The zero-field magnon-phonon detuning decreases progressively with warming, driving the system toward resonance and thereby enhancing their hybridization near the transition temperature. This anomalous thermal evolution provides the crucial key to understanding the unusual terahertz-driven states recently reported in $FePS_3$. More broadly, our results highlight van der Waals magnets as a promising platform for strong coupling among collective excitations and for controlling magnetic states through light-matter interactions.

## I. Introduction

Coupling between collective excitations is central to the emergence and control of complex states in quantum materials. In magnetic systems, interactions between spin and lattice degrees of freedom can hybridize magnons and phonons, producing mixed quasiparticles with properties distinct from those of either excitation alone. Such magnon-phonon hybridization provides a route to transfer energy and angular momentum between different degrees of freedom. It has attracted increasing interest for controlling magnetic states through light-matter interactions. A fundamental question, however, is how such hybridization evolves with temperature. Thermal fluctuations generally broaden quasiparticle excitations and reduce their coherence, suggesting that strong magnon-phonon hybridization should be most robust at low temperatures, where both magnons and phonons are well defined.

Two-dimensional (2D) van der Waals (vdW) magnets provide an interesting platform for investigating such coupled quasiparticles[1-7]. Their reduced dimensionality, strong magnetic anisotropy, and intimate coupling among spin, lattice, charge, and optical degrees of freedom have enabled a range of light-matter phenomena[8-10]. Among them, the vdW antiferromagnet $FePS_3$ is a particularly well-established example of strong spin-lattice coupling[11-20]. $FePS_3$ exhibits Ising-type antiferromagnetism below the Néel temperature ($T_N$~120 K)[2], together with a large magnetic anisotropy[21,22]. Raman and far-infrared spectroscopy have identified low-energy magnons and their strong hybridization with nearby phonons[12-16], including a field-tunable avoided crossing that directly demonstrates strong magnon-phonon coupling. Ultrafast optical and structural probes have further revealed a pronounced coupling between magnetic order and lattice dynamics[17-20].

A recent high-power terahertz experiment[10], however, uncovered an unexpected temperature dependence in this spin-lattice physics. Intense terahertz excitation on the order of 500 MV/cm generates a long-lived metastable magnetic state in $FePS_3$, with a relaxation time of the order of a millisecond. Surprisingly, rather than becoming more stable at low temperatures, the lifetime increases with warming, reaching a maximum near $T_N$. This behavior is completely counterintuitive: increasing temperature normally enhances fluctuations and quasiparticle scattering, conditions expected to weaken, not strengthen coherent hybridization.

Understanding this anomaly therefore requires a full knowledge of how the underlying magnon-phonon hybridization evolves with temperature. Despite extensive spectroscopic studies of $FePS_3$ and direct observation of its field-induced magnon-phonon avoided crossing, the systematic temperature evolution of this system remains unexplored. In particular, it is unknown whether the interaction strength itself changes substantially with temperature, whether the magnon and phonon energies evolve toward or away from resonance, and how these equilibrium properties relate to the unusual temperature dependence observed in the terahertz-driven state.

In this work, we address this fundamental question using temperature-dependent high-field Raman spectroscopy of $FePS_3$. We track the field-induced magnon-phonon avoided crossing from the base temperature to temperatures approaching the antiferromagnetic transition. We find that the avoided-crossing gap, and hence the magnon-phonon coupling strength, remains remarkably robust up to ~ 70 K. In contrast, the zero-field detuning between the magnon and phonon modes progressively decreases with increasing temperature, driving the system toward the resonant strong-hybridization regime. A two-level model quantitatively reproduces this evolution and separates the nearly temperature-independent coupling strength from the strongly temperature-dependent detuning. These results show that thermal evolution does not simply suppress magnon-phonon hybridization in $FePS_3$; instead, warming progressively brings the constituent excitations closer to resonance. This equilibrium behavior provides a spectroscopic counterpart to the anomalous temperature dependence of the terahertz-induced metastable state, highlighting vdW magnets as a fertile platform for strongly coupled spin-lattice quasiparticles.

## II. Experiment

### A. Growth and basic characterization of $FePS_3$ single crystals

We synthesized $FePS_3$ single crystals by chemical vapor transport method. High-purity powders (Thermo Fisher Scientific): iron (99.99%), phosphorus (99.99%), and sulfur (99.998%) were mixed inside an argon-filled glove box. We added an excess of 5 wt% sulfur to the mixture to compensate for its high vapor pressure during growth.

The chemical analysis of the single crystals was performed using a COXEM-EM30 scanning electron microscope equipped with a Bruker QUANTAX 70 energy-dispersive X-ray system to confirm the correct stoichiometry. The *c*-axis spacing was measured by X-ray diffraction using a Rigaku Miniflex II diffractometer. The magnetic properties of the single crystals were measured using a SQUID magnetometer (MPMS, Quantum Design). The room-temperature Raman spectrum of the crystals was measured to identify the Raman peaks.

### B. High magnetic-field Raman spectroscopy measurements

Magneto-Raman measurements were performed in a Physical Property Measurement System (PPMS DynaCool, Quantum Design) using helium as an exchange gas, allowing the sample to be cooled down to 1.8 K under magnetic fields up to 14 T. A 632.8-nm HeNe laser was focused to a ~2-μm spot in backscattering geometry via an objective lens, with incident power maintained at ~400 μW to avoid overheating. The scattered light was dispersed by a Jobin Yvon T64000 triple spectrometer using a 1,800 groove $mm^{-1}$ diffraction grating and detected by a liquid-nitrogen-cooled charge-coupled device.

## III. Results

### A. $FePS_3$ crystals and Raman characterization

$FePS_3$ crystallizes in a monoclinic structure with the space group of $C2/m$, in which Fe atoms form a honeycomb lattice within each vdW layer (Fig. 1a). Below the Néel temperature ($T_N$~ 120 K), the Fe spins form two antiferromagnetic zigzag spin chains with the magnetic easy axis along the out-of-plane ($c$*) direction. The temperature-dependent magnetic susceptibility confirms the antiferromagnetic transition at the Néel temperature $T_N$ of around 120 K (highlighted by the pink dashed line) and the strong magnetic easy axis of $FePS_3$ (Fig. 1b). Additional structural, compositional, and Raman characterization of the crystals is provided in the Supplementary Information (Figs. S1-S3).

To probe the low-energy magnetic excitations and their coupling to phonons, we performed high-field Raman spectroscopy with the magnetic field applied along the $c$* axis (Fig. 1c). Circularly polarized excitation and detection were employed to enhance the relative intensity of the magnetic excitations. This configuration allows the field-dependent magnon modes to be continuously tuned by nearby phonon modes and their hybridization to be directly resolved.

## B. Field-tuned magnon-phonon coupling

We first establish the field-tuned magnon-phonon coupling at the base temperature of 2 K. Figure 2a shows the Raman intensity as a function of the Raman shift and out-of-plane magnetic field. At a base temperature of 2 K (well below $T_N$) and zero field, the low-energy Raman spectrum contains four distinct peaks at around 87.4, 94, 108, and 122 $cm^{-1}$: the latter is a magnon mode, while all three other peaks at lower energy are phonon modes. With increasing magnetic field, the magnon mode splits into two branches (pink for M+, and green for M-) with opposite Zeeman shifts[12] (Fig. 2a-b), whereas the other three phonon modes (P) remain nearly field-independent at low fields. This contrast in field dependence allows magnetic and lattice excitations to be clearly distinguished experimentally. In the Raman intensity map in Fig. 2a, the data points marked by yellow, pink, and green symbols correspond to the positions of the three branches, respectively.

The extracted mode energies are shown in Fig. 2b. As the magnetic field increases, the M- branch blue-shifts linearly. In contrast, the M+ branch red-shifts linearly toward the phonon P peak and the two modes approach resonance near 12 T. Rather than crossing, however, the M+ and P peaks exhibit a definitive avoided crossing, a hallmark of strong magnon-phonon coupling[15]. This behavior is the characteristic signature of strong magnon-phonon coupling in $FePS_3$ and is consistent with previous low-temperature high-field spectroscopic studies.

To quantify the avoided crossing between magnons and phonons, we define the energy separation between the two interacting branches as $\Delta(H)$ (double-headed blue arrow in Fig. 2c). Three quantities characterize the field-tuned coupling: the zero-field detuning ($\Delta_0=\Delta(0)$), the resonance field ($H_c$) at which the two branches reach their minimum separation, and the avoided-crossing gap ($\Delta_{AC}=\Delta(H_c)$). Within a two-level coupled-mode description, the minimum gap is related to the magnon-phonon coupling strength ($g$) by ($\Delta_{AC}=2g$).

The simplified physical picture is illustrated schematically in Fig. 2d. In the absence of coupling, the field-dependent bare magnon mode crosses the nearly field-independent phonon mode at around $H_c$. A finite coupling ($g$) lifts this degeneracy and opens an avoided-crossing gap ($2g$), producing two hybrid magnon-phonon branches. Importantly, the coupling strength ($g$) and the zero-field detuning ($\Delta_0$) represent distinct physical quantities: the former characterizes the interaction between the two excitations, whereas the latter measures their energetic separation before field tuning. Tracking these quantities independently as a function of temperature therefore allows us to determine whether thermal evolution changes the intrinsic coupling strength or instead tunes the two excitations toward or away from resonance.

## C. Anomalous temperature evolution of magnon-phonon coupling

We next investigate how the magnon-phonon coupling evolves with temperature. High-field Raman spectra were measured from 2 to 130 K (see details in Fig. S5-17). As temperature increases, the magnon peaks progressively broaden and become indistinguishable above approximately 90 K, limiting quantitative analysis of the avoided coupling to the ordered state below this temperature. Figure 3a shows the energy separation

*Δ(H)* between the interacting M+ and P peaks as a function of magnetic field at various temperatures.

The temperature evolution of *Δ(H)* reveals two distinct trends. First, the minimum separation between the hybridized modes remains nearly unchanged up to approximately 70 K. Accordingly, the magnon-phonon coupling strength $g=\Delta_{AC}/2$ is remarkably insensitive to temperature over this range (Fig. 3b). At the base temperature, the avoided-crossing gap is approximately 5.7 $cm^{-1}$ ~ 0.71 meV, corresponding to $g$ ~ 2.85 $cm^{-1}$. Despite increasing thermal fluctuations, this strong coupling is not appreciably suppressed on warming.

In contrast, the zero-field detuning $\Delta_0$ decreases progressively with increasing temperature (Fig. 3c). At the same time, the resonance field $H_c$, at which the minimum mode separation occurs, shifts systematically toward a lower magnetic field. The concurrent decreases in $\Delta_0$ and $H_c$ show that, as temperature increases, progressively less magnetic-field tuning is required to bring the magnon and phonon into resonance. Thus, the dominant thermal evolution is not an enhancement of the coupling matrix element itself, but a reduction of the energetic detuning between the two constituent excitations.

This distinction can be quantified by the dimensionless coupling-to-detuning ratio: $\eta(T)=2g(T)/\Delta_0(T)$. As shown in Fig. 3d, $\eta$ increases systematically with temperature. Because $g(T)$ remains nearly constant over the same temperature range, the increase in $\eta$ originates primarily from the decreasing zero-field detuning rather than from an increase in the intrinsic magnon-phonon coupling strength. The equilibrium system therefore evolves progressively toward the resonant strong-hybridization regime as it warms.

This temperature evolution is counterintuitive because increasing thermal fluctuations would broaden quasiparticle excitations, thereby weakening their hybridization. In $FePS_3$, however, the strong magnon-phonon interaction survives while the two modes move closer to resonance. Although magnon broadening above approximately 90 K prevents us from tracking this evolution to $T_N$, the observed trend establishes an unusual thermal pathway toward enhanced magnon-phonon hybridization in the magnetically ordered state.

**IV. Theoretical analysis**

To understand the temperature evolution of the magnon-phonon hybridization, we model the low-energy magnetic excitations of $FePS_3$ using a spin Hamiltonian: $H = J_1 \sum_{\langle i,j\rangle_1} \boldsymbol{S}_1 \cdot \boldsymbol{S}_2 + J_2 \sum_{\langle i,j\rangle_2} \boldsymbol{S}_1 \cdot \boldsymbol{S}_2 + J_3 \sum_{\langle i,j\rangle_3} \boldsymbol{S}_1 \cdot \boldsymbol{S}_2 - K_z \sum_i S_{iz}^2 - g_{\text{eff}} \mu_B B_z \sum_i S_{iz}$ , (1), where $J_1$, $J_2$, and $J_3$ represent the first-, second-, and third-neighbor Heisenberg coupling constants, $K_z$ denotes the single-ion anisotropic interactions, and the last term is the Zeeman energy with $g_{\text{eff}}$ and $B_z$ the effective $g$ factor of $Fe^{2+}$ ions and the strength of the external magnetic field, respectively.

Within linear spin-wave theory[15,21,23] (see Supplemental Note 1), the lowest-energy magnon modes ($M_+$ and $M_-$) are doubly degenerate at zero field ($B_z = 0$) and split under an out-of-plane field because of the Zeeman interaction:

$$E_{\pm}^0 = E_M^0 \mp g_{\text{eff}} \mu_B B_Z \qquad (2)$$

where $E_M^0 = 2S\sqrt{K_z(K_z + J_1 + 4J_2 + 3J_3)}$. The branch approaching the nearby phonon can therefore be represented by a field-dependent bare magnon energy ($E_M(H)$).

We describe its interaction with the phonon using an effective Hamiltonian that is dominated by magnetoelastic coupling due to the exceedingly large single-ion anisotropic interaction[15]:

$$H_{mag-ph} = \begin{pmatrix} \beta_{M_+}^\dagger & b_P^\dagger \end{pmatrix} \begin{pmatrix} E_M^0 - g_{\text{eff}}\mu_B B_z & M_c \\ M_c & E_P^0 \end{pmatrix} \begin{pmatrix} \beta_{M_+} \\ b_P \end{pmatrix}, \qquad (3)$$

where $\beta_{M_+}^\dagger$ and $b_P^\dagger$ denote the creation operators of the magnon mode $M_+$ and phonon mode $P$ whose bare energies are $E_+^0$ and $E_P^0$, respectively, and $M_c$ is the magnon-phonon coupling constant. The experimentally observed magnon splitting constrained the Zeeman slope of the bare magnon, while $E_M^0$=$E_M(0)$, $E_P^0$, and $g$ were obtained by fitting the field-dependent Raman peak positions. As illustrated for 70 K in Fig. 4a, the uncoupled magnon and phonon modes would cross at the resonance field ($H_c$). In contrast, diagonalization of Eq. (3) produces two hybrid branches separated by an avoided-crossing gap $2g$. The calculated eigenvalues reproduce the measured field evolution of the Raman modes over the investigated temperature range.

The fitted bare energies reveal the origin of the unusual temperature dependence observed in Fig. 3. Figure 4b shows $E_M^0(T)$ and $E_P^0(T)$ separately, allowing the temperature evolution of the zero-field detuning, $\Delta_0(T)$=$E_M^0(T) - E_P^0(T)$, to be resolved directly. Their evolution progressively reduces $\Delta_0$ as temperature increases, while the fitted coupling ($g$) remains nearly unchanged over the same temperature range. Thus, the two excitations move closer to resonance primarily because their bare energies shift, not because of a thermally enhanced interaction matrix element.

Fig. 4c illustrates this distinction using the fitted parameters at 2, 40, and 70 K. At 2 K, the bare magnon and phonon are substantially detuned at zero field. As the temperature increases, their energy separation decreases progressively, whereas g remains approximately constant. Temperature therefore moves the zero-field system toward resonance without significantly increasing the intrinsic magnon-phonon coupling. This picture provides a microscopic interpretation of the experimentally observed increase in the coupling-to-detuning ratio ($\eta$=$2g/\Delta_0$) shown in Fig. 3d.

The resulting evolution can also be expressed in terms of the magnon and phonon contents of the hybrid modes. Defining a mixing angle ($\theta$) by $\tan 2\theta = 2g/(E_M^0 - E_P^0)$, the two modes can be written as $\cos\theta\, \beta_{M_+}^\dagger + \sin\theta\, b_P^\dagger$ and $-\sin\theta\, \beta_{M_+}^\dagger + \cos\theta\, b_P^\dagger$. The corresponding magnon and phonon fractions are determined by $\cos^2\theta$ and $\sin^2\theta$. Figure 4d shows that the minority quasiparticle component of the relevant zero-field hybrid mode increases with temperature, approaching a higher degree of mixing as the bare magnon and phonon move toward resonance. The anomalous thermal evolution of $FePS_3$ can therefore be understood as increased magnon–phonon mixing driven primarily by reduced detuning, while the underlying coupling strength remains robust.

## V. Discussions

The present results reveal an unusual evolution of magnon-phonon hybridization in $FePS_3$. The avoided-crossing gap corresponds to a coupling strength of $g$~ 2.8 cm$^{-1}$ (~0.35 meV), which remains nearly unchanged up to approximately 70 K. Together with the large magnetic anisotropy and pronounced spin-lattice responses previously reported in $FePS_3$ [17–22], these observations establish $FePS_3$ as a strongly coupled spin-lattice system.

More importantly, however, the dominant temperature dependence does not originate from a change in the coupling strength itself. Instead, as the system warms, the bare magnon and phonon energies approach one another, reducing their detuning while $g$ remains nearly constant. Consequently, the coupling-to-detuning ratio increases and the zero-field eigenstates acquire increasingly mixed magnon-phonon character. Temperature therefore enhances magnon-phonon hybridization not by strengthening the interaction matrix element, but by driving the constituent excitations toward resonance. This distinction provides a more physical interpretation of the anomalous thermal behavior than attributing it solely to increased coherence.

This equilibrium evolution provides an intriguing connection to the recently reported high-power terahertz-driven metastable state in $FePS_3$[10]. In that experiment, the lifetime of the photoinduced state increases upon warming. It peaks near $T_N$, an unusual behavior attributed to critical magnetic fluctuations and enhanced spin-lattice interactions. Our measurements probe a different regime: Raman spectroscopy characterizes equilibrium magnon-phonon hybridization, whereas intense terahertz excitation creates and follows a nonequilibrium metastable state. Moreover, the increasing magnon linewidth prevents us from quantitatively tracking the avoided crossing well into the $T_N$ phase.

Nevertheless, the two experiments reveal a common and counterintuitive tendency: spin-lattice correlations do not simply weaken upon warming. Instead, our equilibrium measurements show that the magnon and phonon move progressively toward resonance while their coupling remains robust. This temperature-driven enhancement of quasiparticle mixing provides an underlying spectroscopic condition favorable for the unusual nonequilibrium response observed toward the magnetic transition.

The strong coupling observed in $FePS_3$ may also reflect a broader characteristic of vdW magnets. Recent studies have revealed substantial coupling among spin, lattice, electronic, excitonic, and electric-dipole degrees of freedom in several members of this materials family. For example, pressure-dependent studies indicate pronounced exchange striction in several vdW magnets, including materials with substantially larger values than those of $FePS_3$ (Table 1). A dimensionless "exchange striction" coupling constant $\alpha = \frac{d}{\left(\partial d/\partial P\right)} \cdot \frac{\left(\partial T_N/\partial P\right)}{T_N}$ has been widely used to measure the strength of magnon-phonon coupling previously[24,25]. It quantifies, in absolute units, the change in magnetic strength per change in lattice parameter under high pressure, a microscopic definition of spin-lattice coupling. $FePS_3$ has a relatively modest coupling constant of 11, whereas other vdW magnets such as $MnPS_3$, $NiI_2$, and $FeI_2$ exhibit much larger values, even reaching as high as ~175 for $FeI_2$. This compilation of the dimensionless "exchange striction" coupling constant for other vdW magnets hints at an exciting possibility to further explore in the context of understanding and controlling strongly magnon-phonon-coupled systems. Given what has been observed for $FePS_3$, including this work, other vdW magnets with higher $\alpha$ values are likely to host even more gigantic magnon-phonon hybridizations. Although exchange striction cannot be directly related to the spectroscopic coupling $g$ measured here, these observations suggest a broad materials landscape in which strong spin-lattice hybridization may emerge. In this respect, vdW magnets offer a highly tunable platform for exploring and controlling hybrid collective excitations with magnetic fields and light.

In summary, we measured the anomalous temperature dependence of magnon-phonon hybridization in the vdW antiferromagnet $FePS_3$ using high-field temperature-dependent Raman spectroscopy. The magnon-phonon coupling strength detuning progressively decreases, driving the system toward resonance and increasing the mixing of the two quasiparticles. Thus, thermal evolution enhances magnon-phonon hybridization not by increasing the intrinsic coupling strength, but by driving the system toward resonance. This equilibrium behavior provides a spectroscopic counterpart to the unusual temperature dependence of the terahertz-induced metastable state previously observed in $FePS_3$ and points to strong spin-lattice hybridization as a broader, yet novel avenue for controlling collective excitations through light-matter interactions in vdW magnets.

**Data availability**
Data supporting this study are available from the corresponding authors upon request.

**Acknowledgments:** We thank Jihoon Keum for helpful discussions and experimental support. Work at SNU was supported by the Leading Researcher Program of the National Research Foundation of Korea (Grant Nos. 2020R1A3B2079375 and RS-2020-NR049405) and the National Research Foundation (NRF) of Korea (Grant No. RS-2026-25484445), funded by the Ministry of Science and ICT (MSIT). This work was also supported by the National Key Research and Development Program of China (Grant Nos. 2024YFA1408300 and 2022YFA1402704) and the Synergetic Extreme Condition User Facility (SECUF, https://cstr.cn/31123.02.SECUF). The KOMQUEST Center at SNU is supported by the National Research Foundation of Korea (Grant No. RS-2026-25585727).

**Competing interests:** The authors declare no competing interests.

**Additional information:** Supplementary information is available in the online version of this paper. Correspondence and requests for materials should be addressed to the corresponding authors.

**Supplemental Materials include:**
Supplemental Note
Supplemental Figures S1 to S17

## Figures

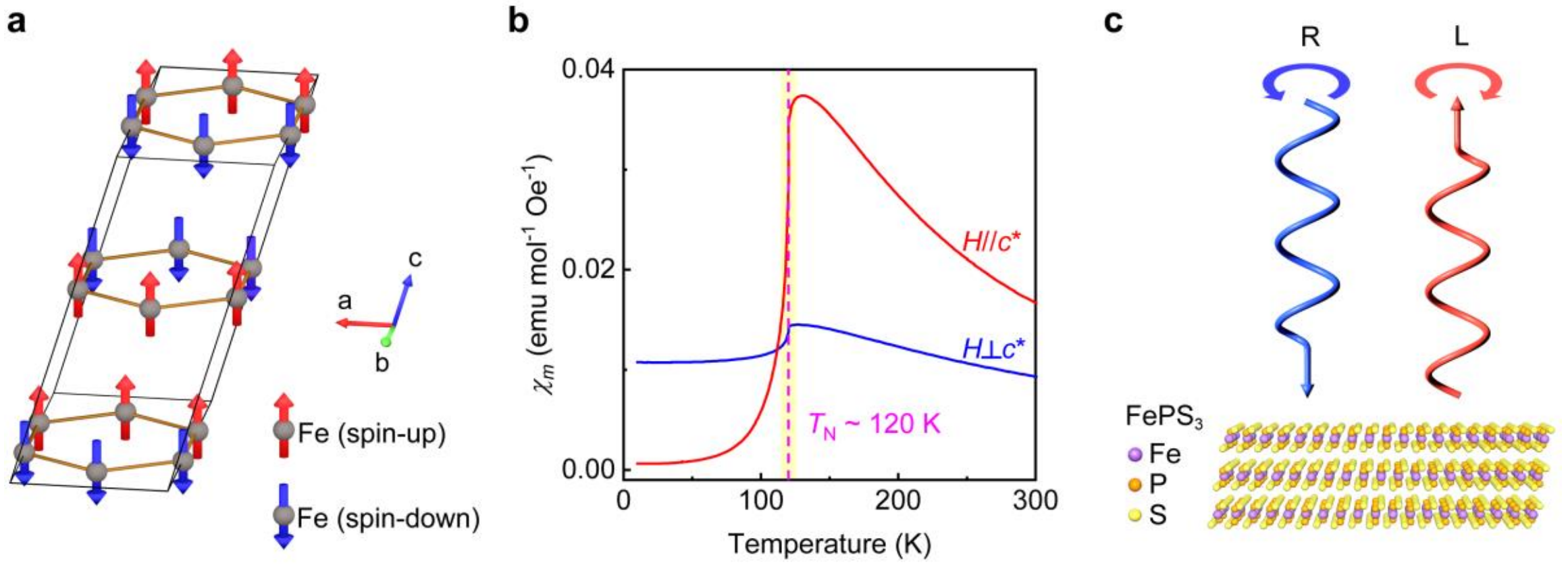

**Fig. 1.** Structural and magnetic properties of high-quality $FePS_3$ crystals. **a**, Schematic of the crystal and magnetic structure of $FePS_3$. **b**, Temperature-dependent magnetic susceptibility with a magnetic field of 3 kOe, applied parallel (red curve) and perpendicular (blue curve) to the $c^*$-axis. The magnetic easy axis is along the $c^*$-axis. The antiferromagnetic transition temperature $T_N$ is around 120 K, indicated by the pink dashed line. **c**, Schematic of the circularly polarized Raman measurements. Right circularly polarized light is normally shone on the sample surface, and the left circularly polarized light is detected correspondingly.

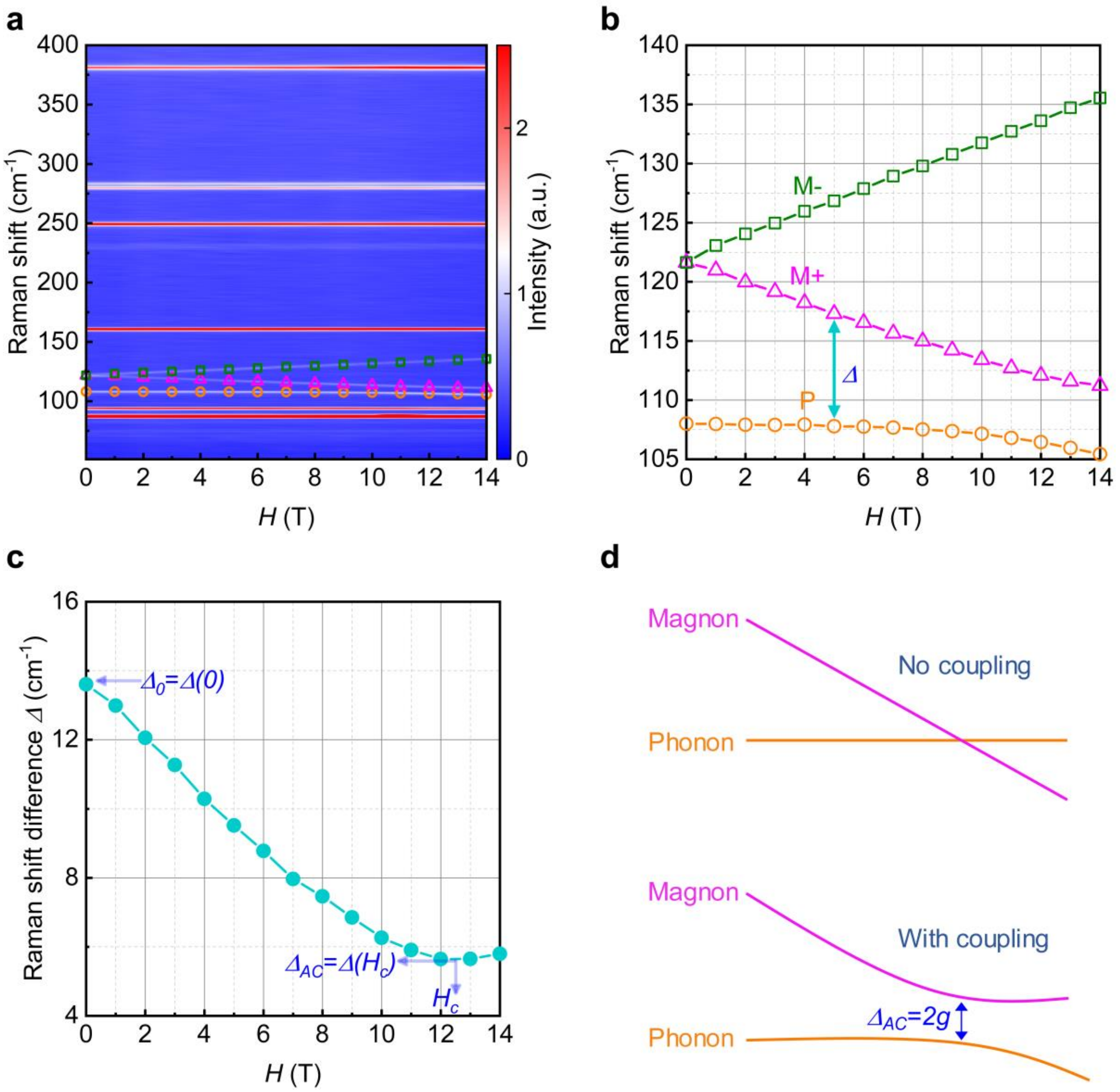

**Fig. 2.** Low-temperature Raman spectrum and magnon-phonon coupling. **a**, Magnetic field and Raman shift mapping of the Raman intensity. Raman peak $P_1$ (in Fig. S3) splits into four peaks at a low temperature of 2 K. As the magnetic field increases, the fourth peak around 122 $cm^{-1}$ splits into two branches (M+ and M-) due to the Zeeman splitting and is thus ascribed to the magnon signal. The neighboring peak around 108 $cm^{-1}$ remains nearly intact and thus is a phonon response (denoted as P). The green, pink, and yellow markers embedded in the contour plot highlight the energy positions of the M-, M+, and P peaks , respectively. **b**, Raman shift versus magnetic field for peaks M- (green curve), M+ (pink curve), and P (yellow curve), similar to that in (**a**). These peak energies are determined by fitting the three peaks as shown in Fig. S4. At a small magnetic field, M+ red-shifts upon increasing the magnetic field, while P remains roughly unchanged. In the high-magnetic-field limit, M+ stays roughly still, but P+ starts to redshift. Such avoided crossing behavior induced by a magnetic field can be regarded as an energy gap opening by the magnon-phonon coupling. To better resolve this physics, $\Delta$ is defined as the Raman shift difference between peaks M+ and P (double-headed blue arrow). **c**, Raman shift difference $\Delta$ versus magnetic field. As the magnetic field increases, $\Delta$ first decreases from the zero-field

detuning $\Delta_0$, then reaches the minimum $\Delta_{AC}$ at the avoided crossing point with critical magnetic field $H_c$, and finally increases after $H_c$, similar to that in (**b**). **d**, Schematic of magnon and phonon branches versus magnetic field without and with quasiparticle coupling. The magnon-phonon coupling would cause the avoided crossing with a gap opening of $\Delta_{AC}=2g$, where $g$ is the magnon-phonon coupling strength.

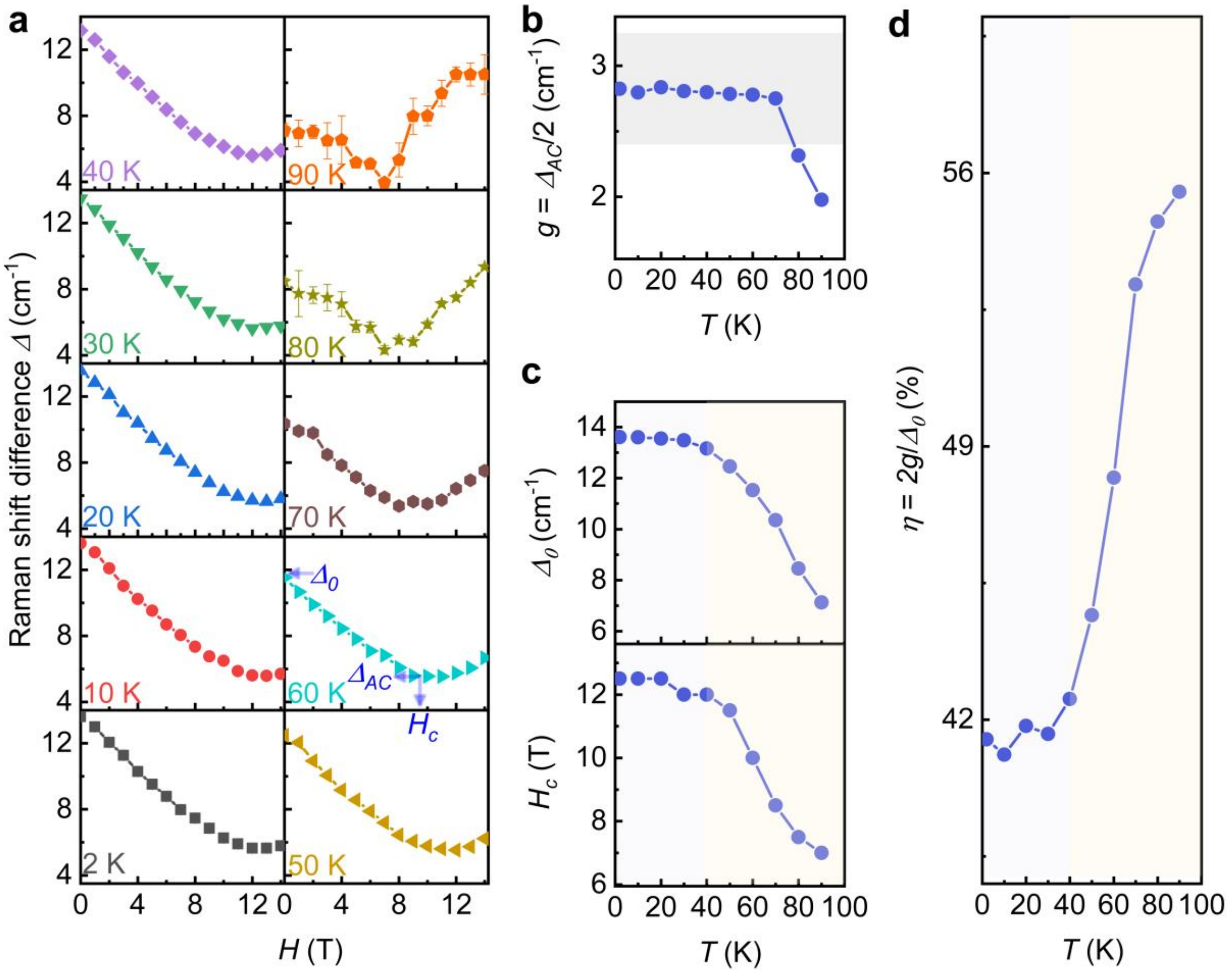

**Fig. 3.** Temperature evolution of the magnon-phonon coupling. **a**, Raman shift difference $\Delta$ versus magnetic field, at 10 different temperatures: 2, 10, 20, 30, 40, 50, 60, 70, 80, and 90 K. From this plot, we can obtain three key parameters: $\Delta_{AC}$, $\Delta_0$, and $H_c$, as represented by the blue arrows, respectively. **b**, Temperature dependence of coupling strength $g=\Delta_{AC}/2$, derived from (**a**). As temperature increases, $\Delta_{AC}$ and the corresponding $g$ remain little changed below 70 K. **c**, Temperature dependence of zero-field detuning $\Delta_0$ and critical field $H_c$, derived from (**a**). As temperature increases, $\Delta_0$ and $H_c$ progressively decrease. **d**, Temperature dependence of the coupling-to-detuning ratio $\eta(T)=2g(T)/\Delta_0(T)$. $\eta$ gradually increases before the magnon peak disappears at 100 K.

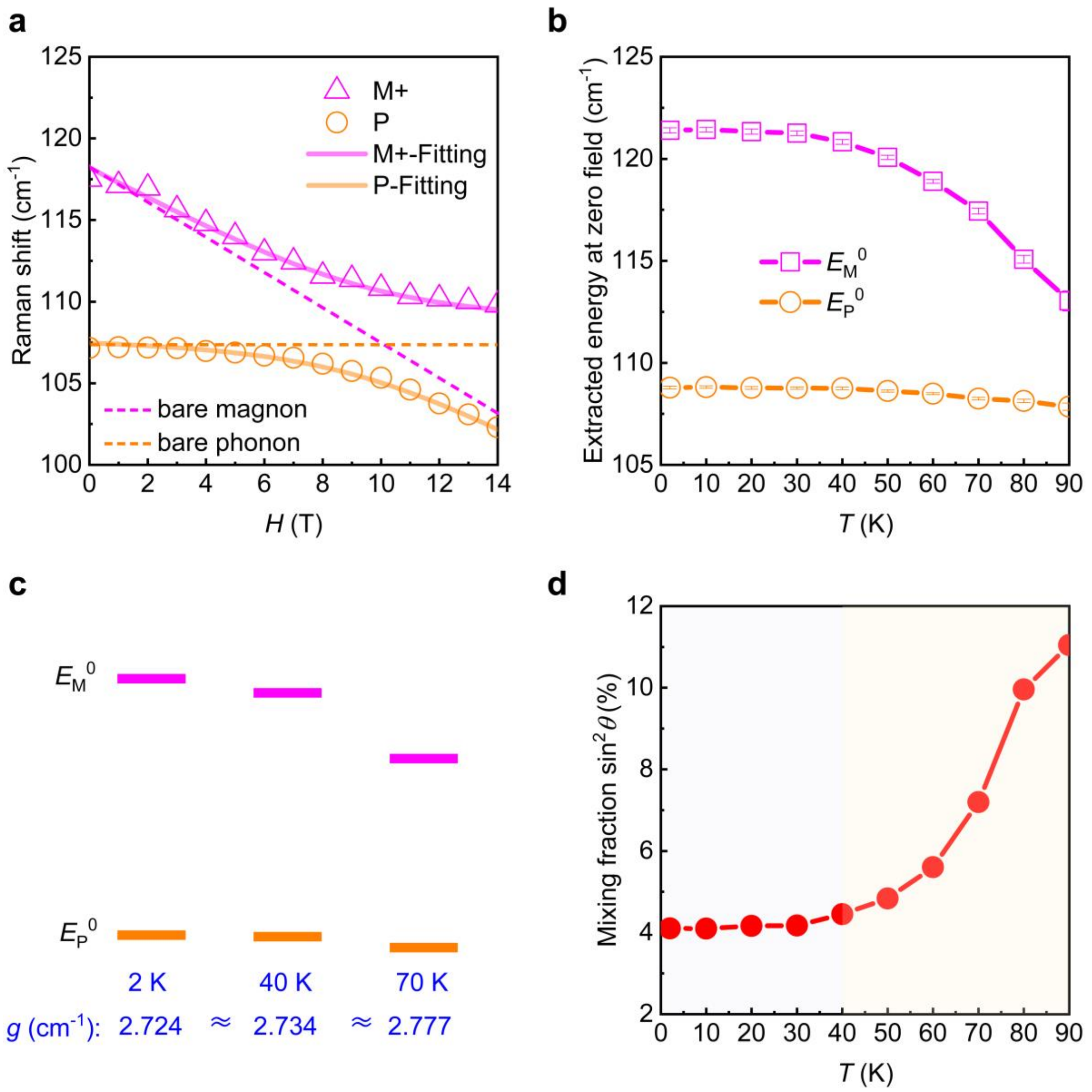

**Fig. 4.** Theoretical modelling of the magnon-phonon coupling. **a**, Raman shift versus magnetic field for the M+ (pink triangle) and P (yellow circle) peaks, taken at 70 K as an example. The pink and yellow curves represent the corresponding fits obtained using the new coupled Hamiltonian in the main text. The dashed lines indicate the field dependence of the bare modes in the absence of coupling. **b**, Extracted bare energies of magnon $E_M{}^0$ (phonon $E_P{}^0$) at zero field versus temperature by fitting. **c**, Energy-level diagrams at 2, 40, and 70 K. The coupling strength $g$ remains little changed, while rising temperature brings the bare modes closer to resonance. **d**, Temperature dependence of the mixing fraction $\sin^2\theta$, where the mixing angle $\theta$ is defined as $\tan 2\theta = 2g/(E_M^0 - E_P^0)$. The mixing fraction increases with temperature.

| | $\alpha = \frac{d}{\left(\partial d/_{\partial P}\right)} \cdot \frac{\left(\partial T_N/_{\partial P}\right)}{T_N}$ | **Reference** |
|---|---|---|
| $FePS_3$ | $-11.09$ | Ref. [26,27] |
| $MnPS_3$ | $-47.39$ | Ref. [28] |
| $NiPS_3$ | $3.31 \sim 5.70$ | Ref. [29] |
| $NiPSe_3$ | $4.17 \sim 12.11$ | Ref. [30] |
| $CrPS_4$ | $5.36 \sim 20.12$ | Ref. [31] |
| $Cr_2Ge_2Te_6$ | $16.99 \sim 59.45$ | Ref. [32,33] |
| $NiI_2$ | $-48.55 \sim -47.44$ | Ref. [34] |
| $FeI_2$ | $-175.07$ | Ref. [35,36] |

**Table 1.** Summary of exchange-striction parameters that are estimated using the data in the references, as given for selected vdW magnets.

# Supplemental Material for:

## Thermally enhanced magnon-phonon resonance in van der Waals antiferromagnet $FePS_3$

Kai-Xuan Zhang[1,4†*], Beom Hyun Kim[1†], Seungbok Lee[1], Hao Li[2], Eugene Park[1$], Heejun Yang[1,3], Feng Jin[2], Qingming Zhang[2*], and Je-Geun Park[1,4*]

[1]*Department of Physics and Astronomy, Seoul National University, Seoul 08826, South Korea*
[2]*Beijing National Laboratory for Condensed Matter Physics, Institute of Physics, Chinese Academy of Sciences, Beijing 100190, China*
[3]*RIKEN Center for Emergent Matter Science (CEMS), Wako, Saitama 351-0198, Japan*
[4]*KOMQUEST Center, Seoul National University, Seoul 08826, South Korea*

[$]Present address: Department of Physics, Massachusetts Institute of Technology, Cambridge, MA 02139, USA

[†] These authors contributed equally to this work.
[*]Corresponding authors: Kai-Xuan Zhang (email: kxzhang.research@gmail.com), Qingming Zhang (email: qmzhang@iphy.ac.cn), and Je-Geun Park (email: jgpark10@snu.ac.kr)

**This PDF file includes:**

Supplemental Note

Supplemental Figures S1 to S17

**Supplemental Note 1. Linear spin-wave theory**

The magnetic Hamiltonian of $FePS_3$ in Eq. (1) is rewritten as

$$H=\frac{1}{2}\sum_j\sum_{i=1}^4\left[J_1\sum_{k=1}^3\boldsymbol{S}_i(\boldsymbol{R}_j)\cdot\boldsymbol{S}_{i_{k_1}}\left(\boldsymbol{R}_{j_{k_1}}\right)+J_2\sum_{k=1}^6\boldsymbol{S}_j(\boldsymbol{R})\cdot\boldsymbol{S}_{i_{k_2}}\left(\boldsymbol{R}_{j_{k_2}}\right)\right]$$
$$+\frac{1}{2}\sum_j\sum_{i=1}^4\left[J_3\sum_{k=1}^3\boldsymbol{S}_i(\boldsymbol{R}_j)\cdot\boldsymbol{S}_{i_{k_3}}\left(\boldsymbol{R}_{j_{k_3}}\right)\right]-K_z\sum_j\sum_{i=4}^4 S_{iz}^2(\boldsymbol{R}_j)$$
$$-g_{\mathrm{eff}}\mu_B B_z\sum_j\sum_{i=1}^4 S_{iz}(\boldsymbol{R}_j), \tag{S1}$$

where $\boldsymbol{S}_i(\boldsymbol{R}_j)$ denotes the spin operator at the $i$th basis site in the $j$th unit cell whose lattice vector is $\boldsymbol{R}_j$, and $i_{k_1}$ and $j_{k_1}$ ($i_{k_2}$ and $j_{k_2}$, $i_{k_3}$ and $j_{k_3}$) represent the basis and unit-cell indices of its first (second, third) nearest neighboring spins, respectively. Note that $FePS_3$ exhibits the zigzag magnetic order in which the spin direction is parallel to the $c^*(z)$ axis. Only $S_{iz}(\boldsymbol{R}_j)$ spin operators have non-zero expectation values as $\langle S_{1z}(\boldsymbol{R}_j)\rangle=-\langle S_{2z}(\boldsymbol{R}_j)\rangle=-\langle S_{3z}(\boldsymbol{R}_j)\rangle=\langle S_{4z}(\boldsymbol{R}_j)\rangle=S$. We can perform the Holstein-Primakoff transformations as follows:

$$S_{1,+}(\boldsymbol{R}_j)=\sqrt{2S}\;a_{1j},\quad S_{1,-}(\boldsymbol{R}_j)=\sqrt{2S}\;a_{1j}^\dagger,\quad S_{1,z}(\boldsymbol{R}_j)=S-a_{1j}^\dagger a_{1j}, \tag{S2.1}$$
$$S_{2,+}(\boldsymbol{R}_j)=\sqrt{2S}\;b_{1j}^\dagger,\quad S_{2,-}(\boldsymbol{R}_j)=\sqrt{2S}\;b_{1j},\quad S_{2,z}(\boldsymbol{R}_j)=-S+b_{1j}^\dagger b_{1j}, \tag{S2.2}$$
$$S_{3,+}(\boldsymbol{R}_j)=\sqrt{2S}\;b_{2j}^\dagger,\quad S_{3,-}(\boldsymbol{R}_j)=\sqrt{2S}\;b_{2j},\quad S_{3,z}(\boldsymbol{R}_j)=-S+b_{2j}^\dagger b_{2j}, \tag{S2.3}$$
$$S_{4,+}(\boldsymbol{R}_j)=\sqrt{2S}\;a_{2j},\quad S_{4,-}(\boldsymbol{R}_j)=\sqrt{2S}\;a_{2j}^\dagger,\quad S_{4,z}(\boldsymbol{R}_j)=S-a_{2j}^\dagger a_{2j}, \tag{S2.4}$$

where $a_{1j}^\dagger$, $b_{1j}^\dagger$, $b_{2j}^\dagger$, and $a_{2j}^\dagger$ ($a_{1j}$ $b_{1j}$, $b_{2j}$, and $a_{2j}$) are the creation (annihilation) operators of Holstein-Primakoff bosons (magnons) at 1th- to 4th-basis sites in the $j$th unit cell, respectively.

According to the linear spin-wave theory, we can deduce the magnon Hamiltonian as follows:

$$H=2NS(S+1)(J_1-2J_2-3J_3-2K_z)-4SJ_2\sum_{\boldsymbol{k}}\cos\boldsymbol{k}\cdot\boldsymbol{a}$$
$$+\frac{S}{2}\sum_{\boldsymbol{k}}\begin{pmatrix}b_{1\boldsymbol{k}}^\dagger & b_{2\boldsymbol{k}}^\dagger & a_{1,-\boldsymbol{k}} & a_{2,-\boldsymbol{k}}\end{pmatrix}\begin{pmatrix}A_2(\boldsymbol{k}) & B(\boldsymbol{k}) & C^*(\boldsymbol{k}) & F^*(\boldsymbol{k})\\ B^*(\boldsymbol{k}) & A_2(\boldsymbol{k}) & D^*(\boldsymbol{k}) & E^*(\boldsymbol{k})\\ C(\boldsymbol{k}) & D(\boldsymbol{k}) & A_1(\boldsymbol{k}) & B^*(\boldsymbol{k})\\ F(\boldsymbol{k}) & E(\boldsymbol{k}) & B(\boldsymbol{k}) & A_1(\boldsymbol{k})\end{pmatrix}\begin{pmatrix}b_{1\boldsymbol{k}}\\ b_{2\boldsymbol{k}}\\ a_{1,-\boldsymbol{k}}^\dagger\\ a_{2,-\boldsymbol{k}}^\dagger\end{pmatrix}$$
$$+\frac{S}{2}\sum_{\boldsymbol{k}}\begin{pmatrix}a_{1\boldsymbol{k}}^\dagger & a_{2\boldsymbol{k}}^\dagger & b_{1,-\boldsymbol{k}} & b_{2,-\boldsymbol{k}}\end{pmatrix}\begin{pmatrix}A_1(\boldsymbol{k}) & B(\boldsymbol{k}) & C(\boldsymbol{k}) & D(\boldsymbol{k})\\ B^*(\boldsymbol{k}) & A_1(\boldsymbol{k}) & F(\boldsymbol{k}) & E(\boldsymbol{k})\\ C^*(\boldsymbol{k}) & F^*(\boldsymbol{k}) & A_2(\boldsymbol{k}) & B^*(\boldsymbol{k})\\ D^*(\boldsymbol{k}) & E^*(\boldsymbol{k}) & B(\boldsymbol{k}) & A_2(\boldsymbol{k})\end{pmatrix}\begin{pmatrix}a_{1\boldsymbol{k}}\\ a_{2\boldsymbol{k}}\\ b_{1,-\boldsymbol{k}}^\dagger\\ b_{2,-\boldsymbol{k}}^\dagger\end{pmatrix}, \tag{S3}$$

where $N$ denotes the total number of unit cell, and $a_{1\boldsymbol{k}}^\dagger$, $a_{2\boldsymbol{k}}^\dagger$, $b_{1\boldsymbol{k}}^\dagger$, and $b_{2\boldsymbol{k}}^\dagger$ are the Fourier transformation of Hostein-Primakoff bosons $a_{1j}^\dagger$, $b_{1j}^\dagger$, $b_{2j}^\dagger$, and $a_{2j}^\dagger$, respectively.

Here, $A_1(\boldsymbol{k})=-J_1+2J_2+3J_3+2K_z+\frac{g_{\mathrm{eff}}\mu_B B_z}{S}+2J_2\cos\boldsymbol{k}\cdot\boldsymbol{a}$,
$A_2(\boldsymbol{k})=-J_1+2J_2+3J_3+2K_z-\frac{g_{\mathrm{eff}}\mu_B B_z}{S}+2J_2\cos\boldsymbol{k}\cdot\boldsymbol{a}$, $B(\boldsymbol{k})=J_1\left(1+e^{i\boldsymbol{k}\cdot\boldsymbol{a}}\right)$,
$C(\boldsymbol{k})=J_1+2J_3\cos\boldsymbol{k}\cdot\boldsymbol{a}+J_3e^{i\boldsymbol{k}\cdot\boldsymbol{b}}$, $E(\boldsymbol{k})=J_1e^{i\boldsymbol{k}\cdot\boldsymbol{b}}+2J_3e^{i\boldsymbol{k}\cdot\boldsymbol{b}}\cos\boldsymbol{k}\cdot\boldsymbol{a}+J_3$,
$D(\boldsymbol{k})=J_2\left(1+e^{i\boldsymbol{k}\cdot\boldsymbol{a}}\right)\left(1+e^{i\boldsymbol{k}\cdot\boldsymbol{b}}\right)$, $F(\boldsymbol{k})=J_2\left(1+e^{-i\boldsymbol{k}\cdot\boldsymbol{a}}\right)\left(1+e^{i\boldsymbol{k}\cdot\boldsymbol{b}}\right)$,
where $\boldsymbol{a}$ and $\boldsymbol{b}$ denote two primitive vectors parallel to the $ab$ plane. At $\Gamma$ point ($\boldsymbol{k}=$

$\mathbf{0}$), we obtained two separate Hamiltonian $H_+$ and $H_-$ corresponding to two magnon branches determined by the ground spin directions as follows:

$$H_+ = \frac{S}{2}\begin{pmatrix} a_{1\Gamma}^\dagger & a_{2\Gamma}^\dagger & b_{1\Gamma} & b_{2\Gamma} \end{pmatrix} \begin{pmatrix} A_1 & B & C & D \\ B & A_1 & D & C \\ C & D & A_2 & B \\ D & C & B & A_2 \end{pmatrix} \begin{pmatrix} a_{1\Gamma} \\ a_{2\Gamma} \\ b_{1\Gamma}^\dagger \\ b_{2\Gamma}^\dagger \end{pmatrix}, \qquad \text{(S4)}$$

$$H_- = \frac{S}{2}\begin{pmatrix} b_{1\Gamma}^\dagger & b_{2\Gamma}^\dagger & a_{1\Gamma} & a_{2\Gamma} \end{pmatrix} \begin{pmatrix} A_2 & B & C & D \\ B & A_2 & D & C \\ C & D & A_1 & B \\ D & C & B & A_1 \end{pmatrix} \begin{pmatrix} b_{1\Gamma} \\ b_{2\Gamma} \\ a_{1\Gamma}^\dagger \\ a_{2\Gamma}^\dagger \end{pmatrix}, \qquad \text{(S5)}$$

where

$$A_1 = A + \frac{g_{\text{eff}}\mu_B B_z}{S},\ A_2 = A - \frac{g_{\text{eff}}\mu_B B_z}{S}\ ,\ A = -J_1 + 4J_2 + 3J_3 + 2K_z,$$
$$B = 2J_1, \quad C = J_1 + 3J_3, \quad D(\boldsymbol{k}) = 4J_2.$$

Using Eq. (S4) and (S5), we obtained the dynamical equation of motions as follows:

$$i\hbar\partial_t \begin{pmatrix} a_{1\Gamma} \\ a_{2\Gamma} \\ b_{1\Gamma}^\dagger \\ b_{2\Gamma}^\dagger \end{pmatrix} = \hbar\omega_+ \begin{pmatrix} a_{1\Gamma} \\ a_{2\Gamma} \\ b_{1\Gamma}^\dagger \\ b_{2\Gamma}^\dagger \end{pmatrix} = \begin{pmatrix} A_1 & B & C & D \\ B & A_1 & D & C \\ -C & -D & -A_2 & -B \\ -D & -C & -B & -A_2 \end{pmatrix} \begin{pmatrix} a_{1\Gamma} \\ a_{2\Gamma} \\ b_{1\Gamma}^\dagger \\ b_{2\Gamma}^\dagger \end{pmatrix}, \qquad \text{(S6)}$$

$$i\hbar\partial_t \begin{pmatrix} b_{1\Gamma} \\ b_{2\Gamma} \\ a_{1\Gamma}^\dagger \\ a_{2\Gamma}^\dagger \end{pmatrix} = \hbar\omega_- \begin{pmatrix} b_{1\Gamma} \\ b_{2\Gamma} \\ a_{1\Gamma}^\dagger \\ a_{2\Gamma}^\dagger \end{pmatrix} = \begin{pmatrix} A_2 & B & C & D \\ B & A_2 & D & C \\ -C & -D & -A_1 & -B \\ -D & -C & -B & -A_1 \end{pmatrix} \begin{pmatrix} b_{1\Gamma} \\ b_{2\Gamma} \\ a_{1\Gamma}^\dagger \\ a_{2\Gamma}^\dagger \end{pmatrix}. \qquad \text{(S7)}$$

Eigenvalues of Eq. (S6) and (S7) are presented as:

$$\hbar\omega_+ = E_M^0 - g_{\text{eff}}\mu_B B_z, \quad E_M' - g_{\text{eff}}\mu_B B_z$$
$$\hbar\omega_- = E_M^0 + g_{\text{eff}}\mu_B B_z, \quad E_M' + g_{\text{eff}}\mu_B B_z, \qquad \text{(S8)}$$

where

$$E_M^0 = S\sqrt{(A+B)^2 - (C+D)^2} = 2S\sqrt{K_z(J_1 + 4J_2 + 3J_3 + K_z)}$$
$$E_M' = S\sqrt{(A-B)^2 - (C-D)^2} = 2S\sqrt{(-J_1 + 3J_3 + K_z)(-2J_1 + 4J_2 + K_z)}.$$

Therefore, the energy of the two lowest magnon modes are given as:

$$E_\pm^0 = E_M^0 \mp g_{\text{eff}}\mu_B B_z. \qquad \text{(S9)}$$

**Supplemental Figures**

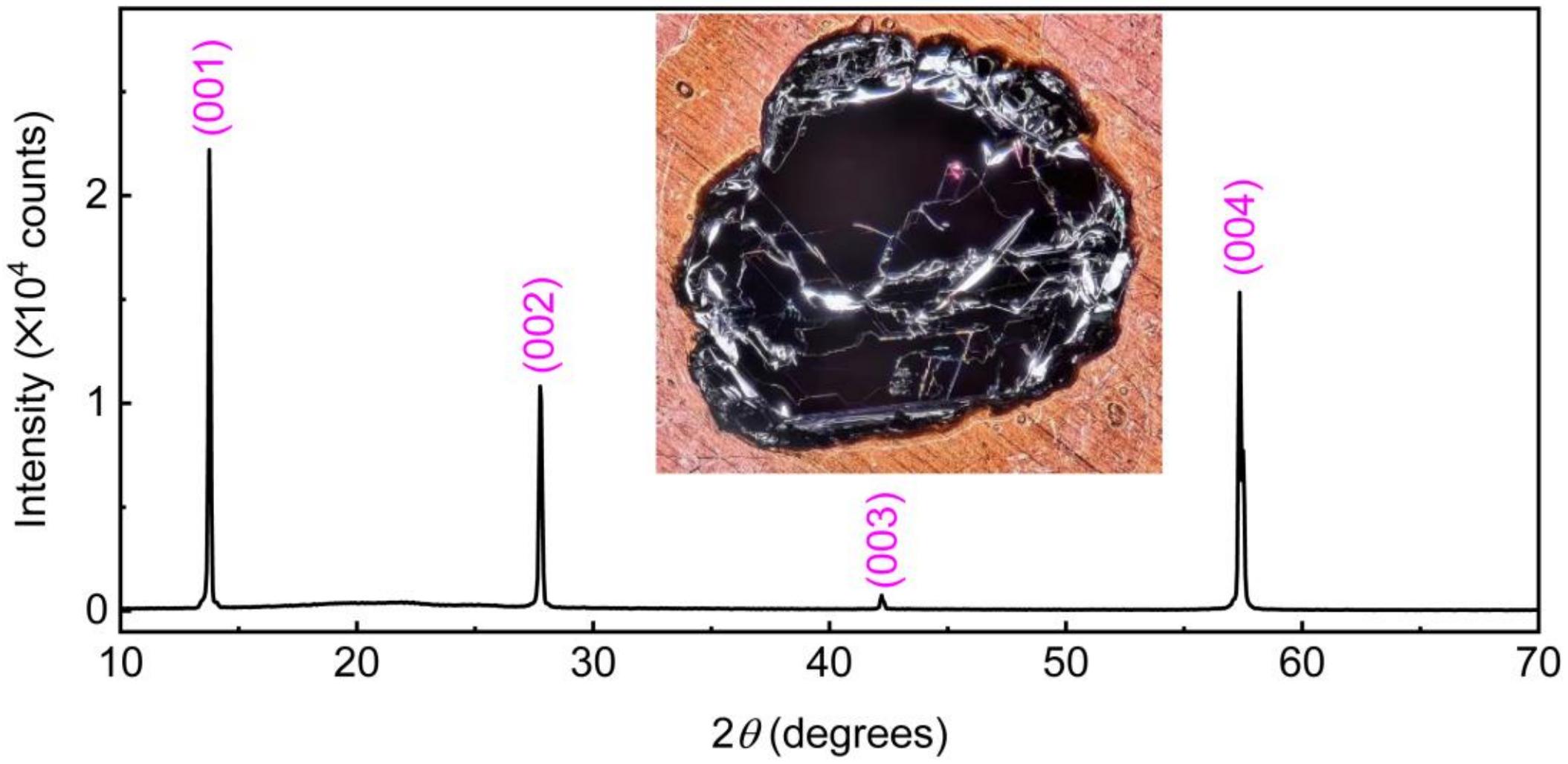

**Fig. S1.** X-ray diffraction (XRD) pattern of $FePS_3$ single crystals. The XRD pattern exhibits well-resolved, sharp Bragg peaks of the (001) facets, confirming high crystallinity.

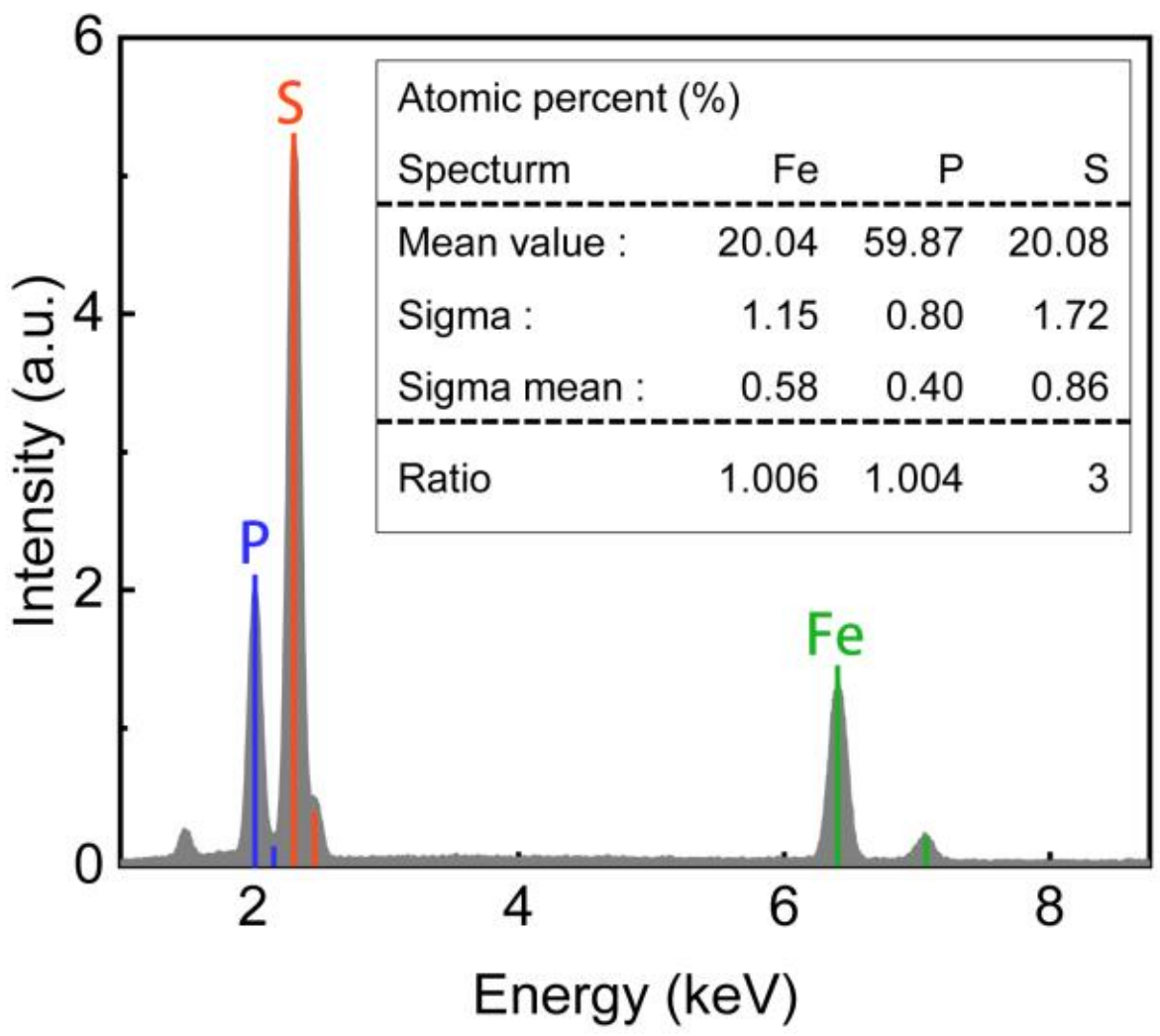

**Fig. S2.** Energy-dispersive spectroscopy (EDS) of $FePS_3$ single crystals. The EDS spectrum yields an element ratio of Fe:P:S ≈ 1.006:1.004:3, consistent with the stoichiometric ratio of 1:1:3 and thus confirming the stoichiometric quality.

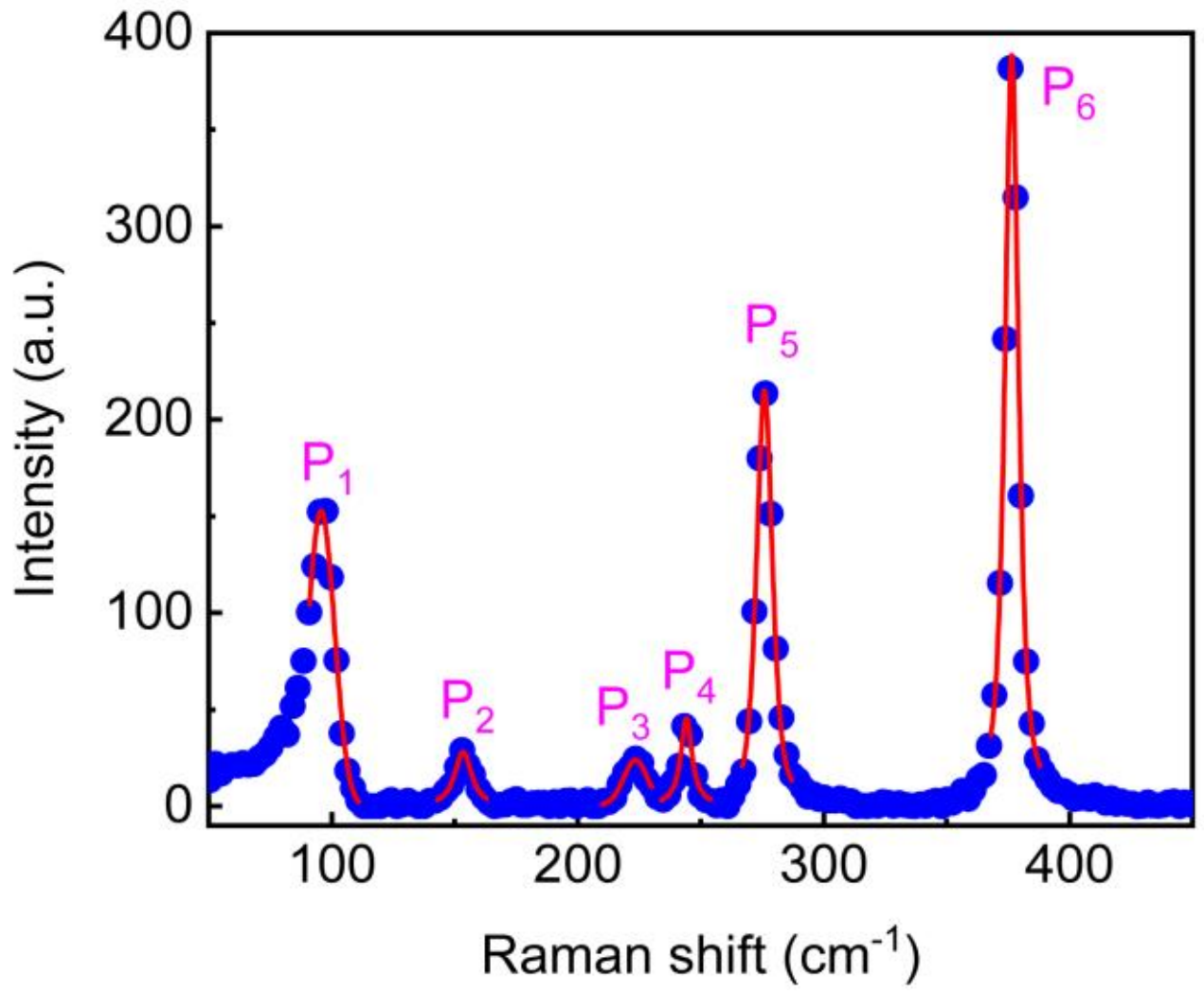

**Fig. S3.** Room-temperature Raman spectrum of $FePS_3$. Six Raman peaks ($P_1$-$P_6$) are highlighted in red, with the wavenumbers at 95.7, 153.5, 223.5, 244.2, 276.0, and 376.5 $cm^{-1}$, which are in excellent agreement with the previous report[2].

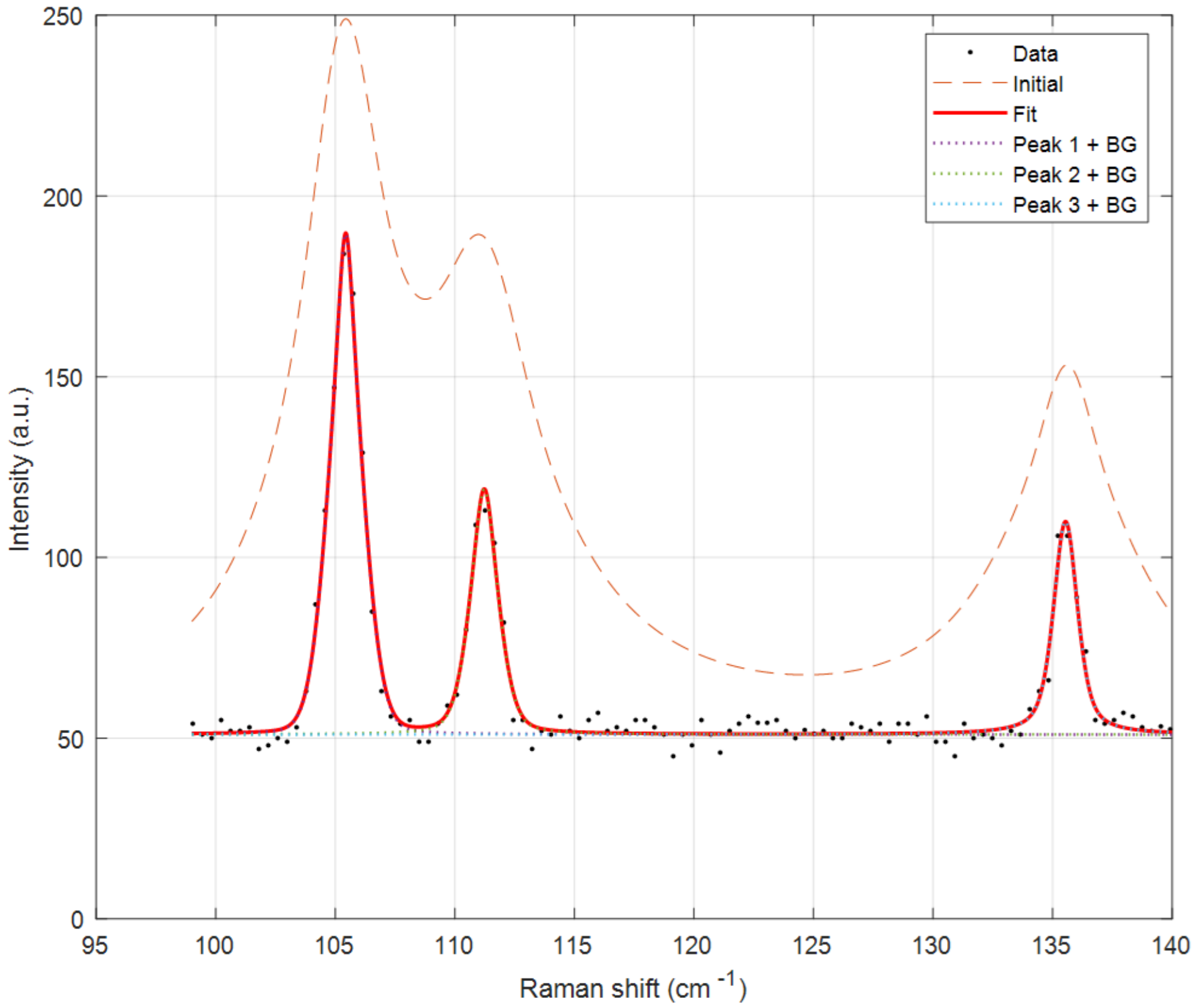

**Fig. S4.** Fitting example for Raman spectrum under 2 K and 14 T. The three peaks (P, M+, and M-) are well fitted with minimal error bars by the Voigt functions (red curves).

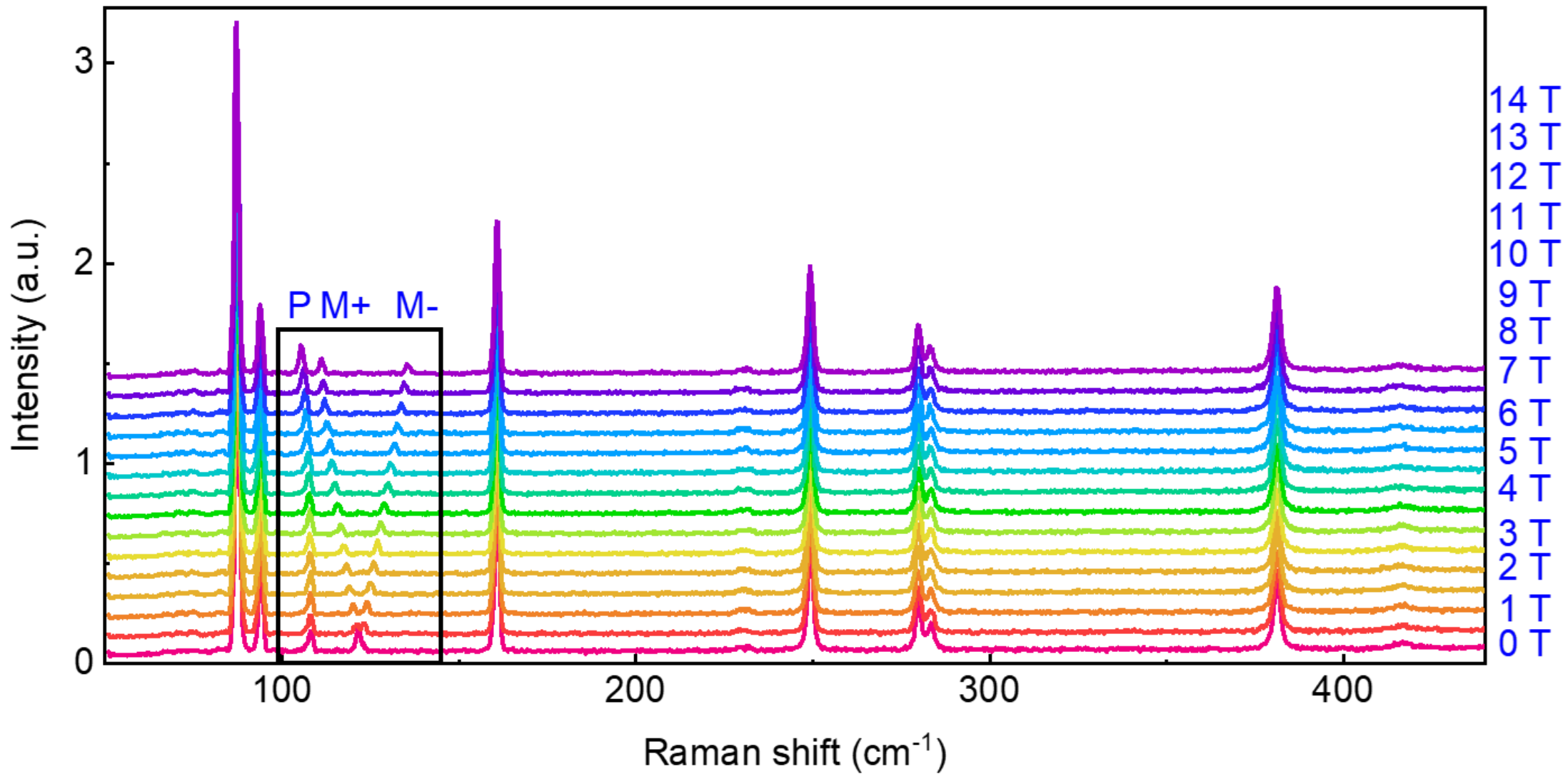

**Fig. S5.** Raw Raman spectra at 10 K under various magnetic fields. The black box indicates the energy range of interest. The P, M+, and M- peaks represent the phonon and split magnon modes under an applied magnetic field.

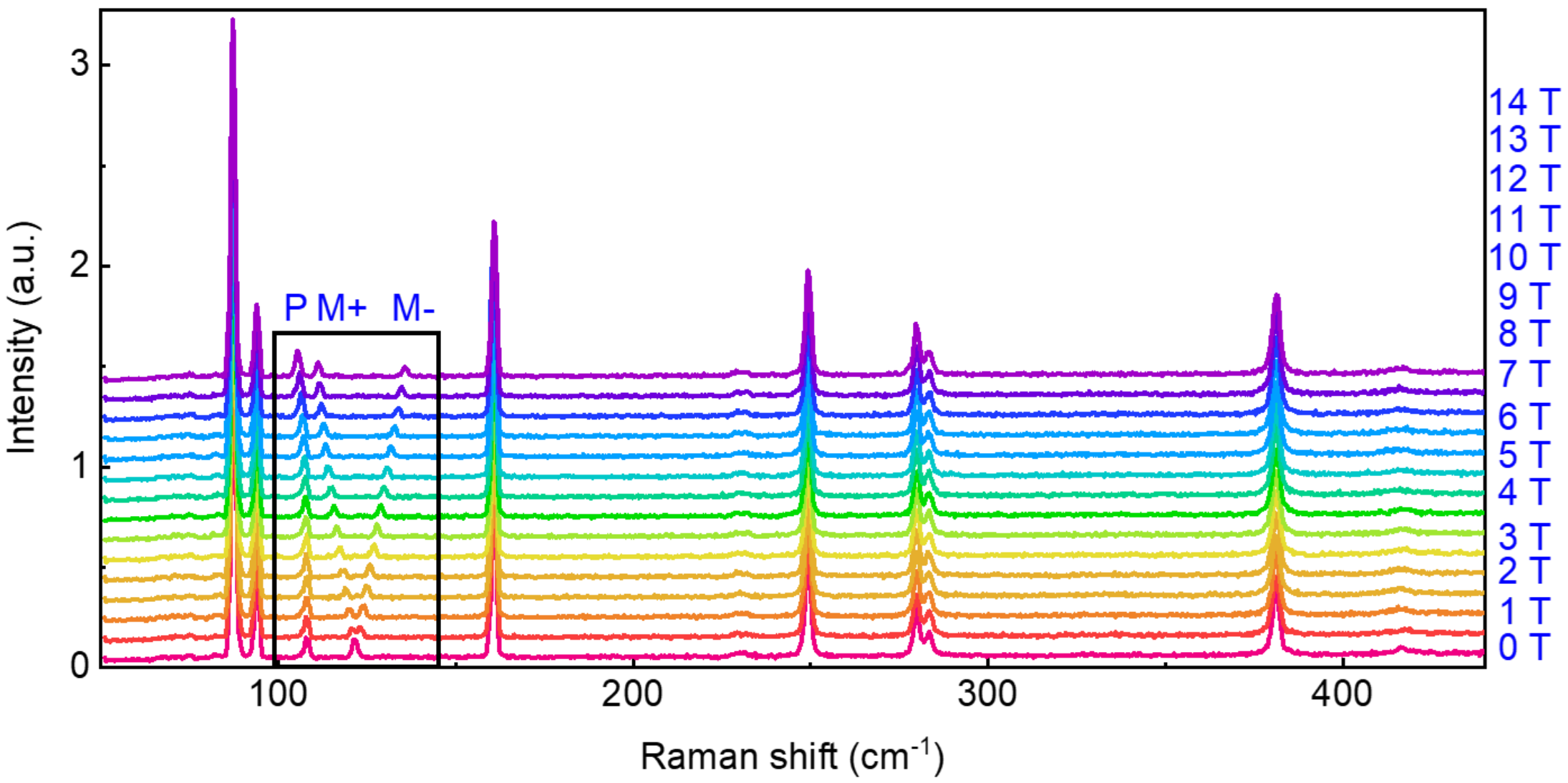

**Fig. S6.** Raw Raman spectra at 20 K under various magnetic fields. The black box indicates the energy range of interest. The P, M+, and M- peaks represent the phonon and split magnon modes under an applied magnetic field.

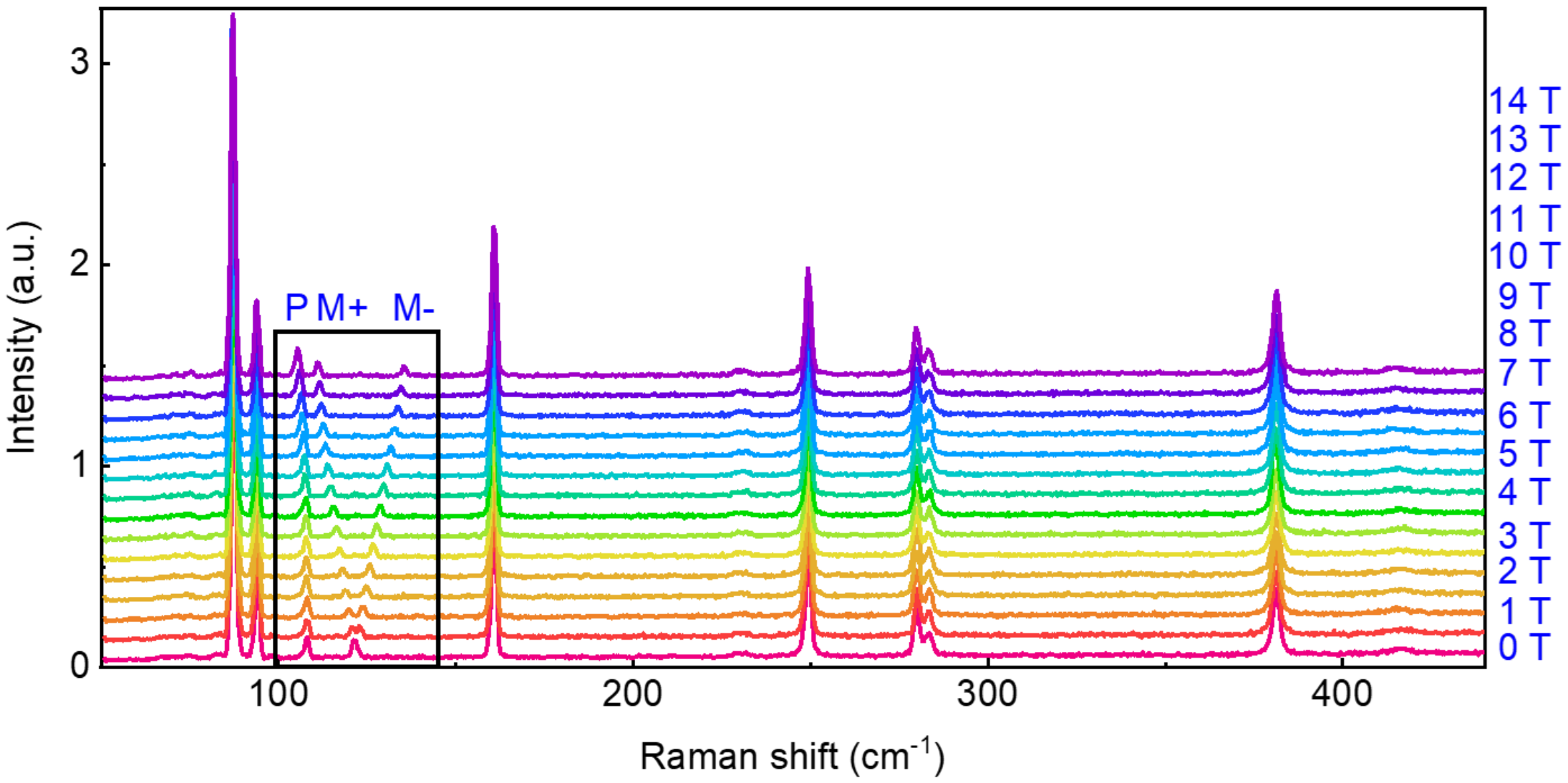

**Fig. S7.** Raw Raman spectra at 30 K under various magnetic fields. The black box indicates the energy range of interest. The P, M+, and M- peaks represent the phonon and split magnon modes under an applied magnetic field.

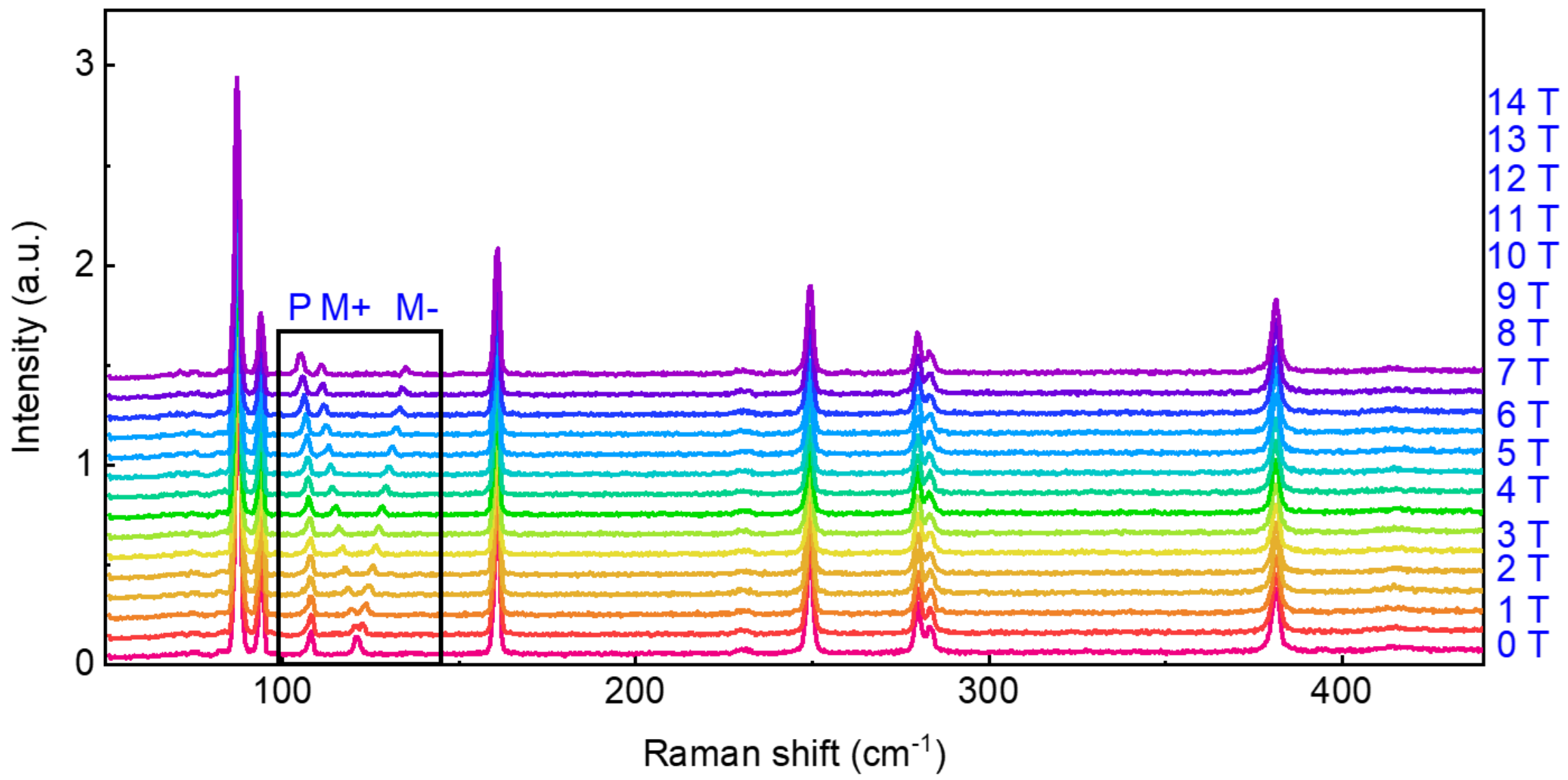

**Fig. S8.** Raw Raman spectra at 40 K under various magnetic fields. The black box indicates the energy range of interest. The P, M+, and M- peaks represent the phonon and split magnon modes under an applied magnetic field.

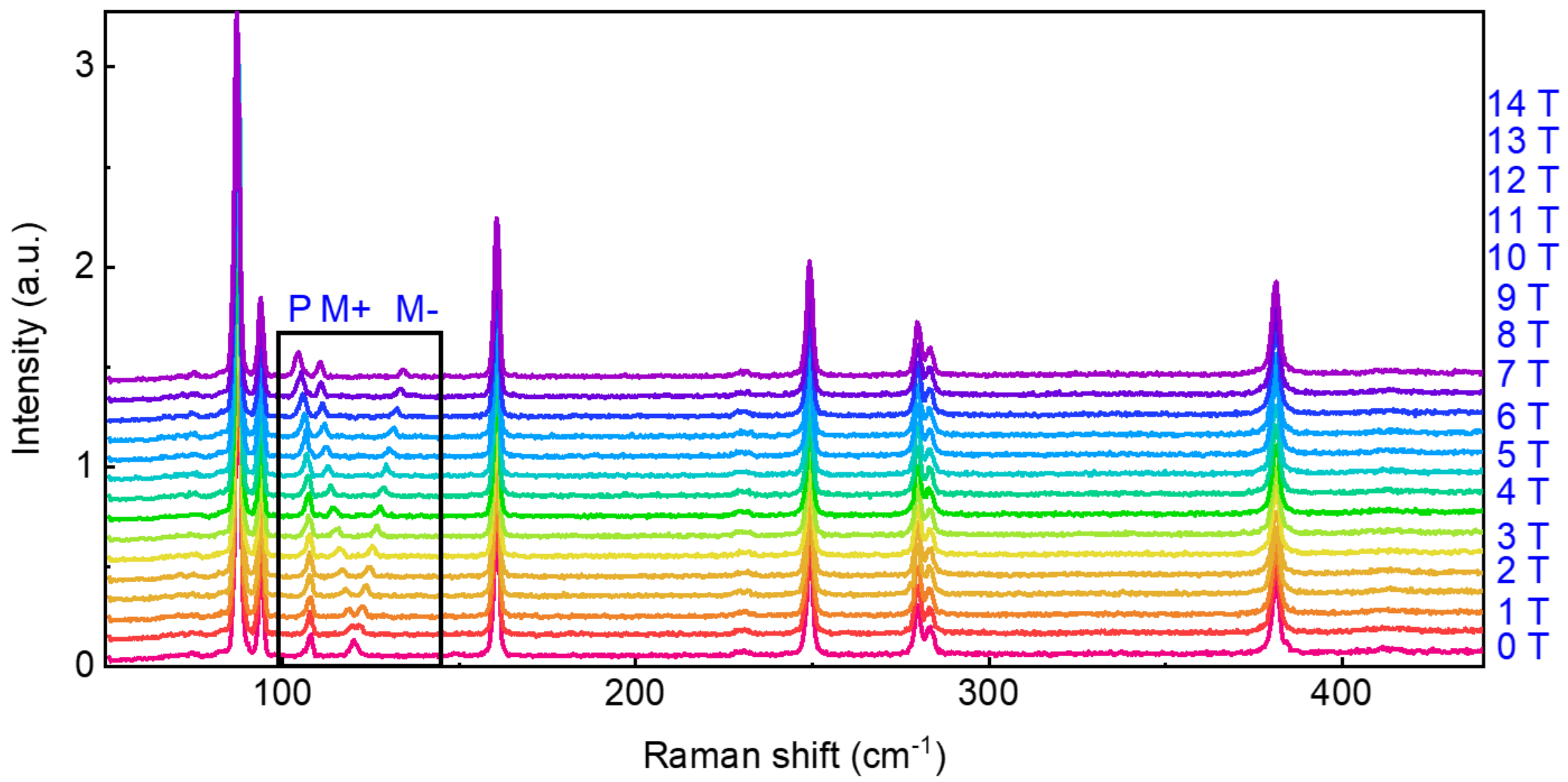

**Fig. S9.** Raw Raman spectra at 50 K under various magnetic fields. The black box indicates the energy range of interest. The P, M+, and M- peaks represent the phonon and split magnon modes under an applied magnetic field.

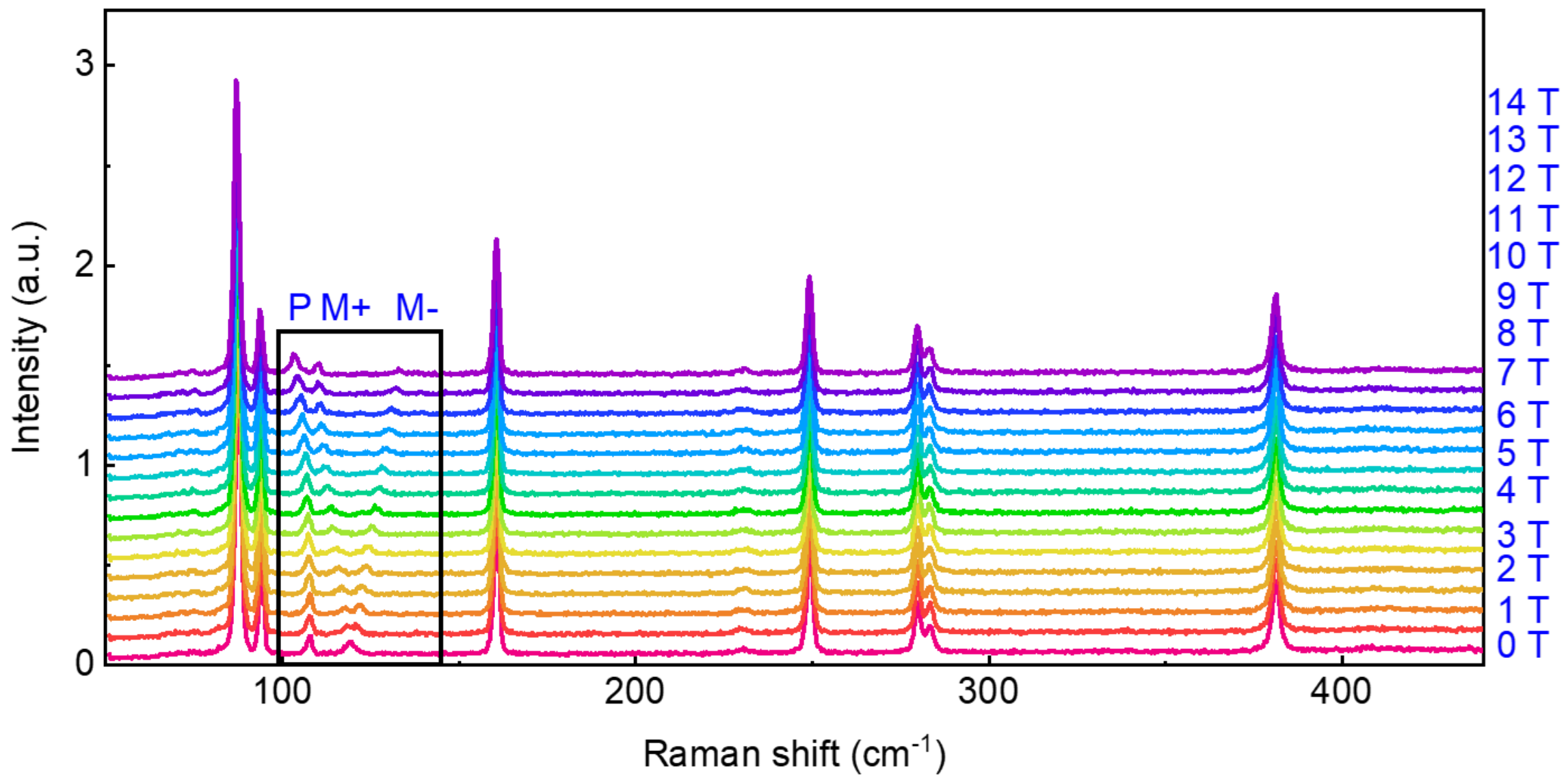

**Fig. S10.** Raw Raman spectra at 60 K under various magnetic fields. The black box indicates the energy range of interest. The P, M+, and M- peaks represent the phonon and split magnon modes under an applied magnetic field.

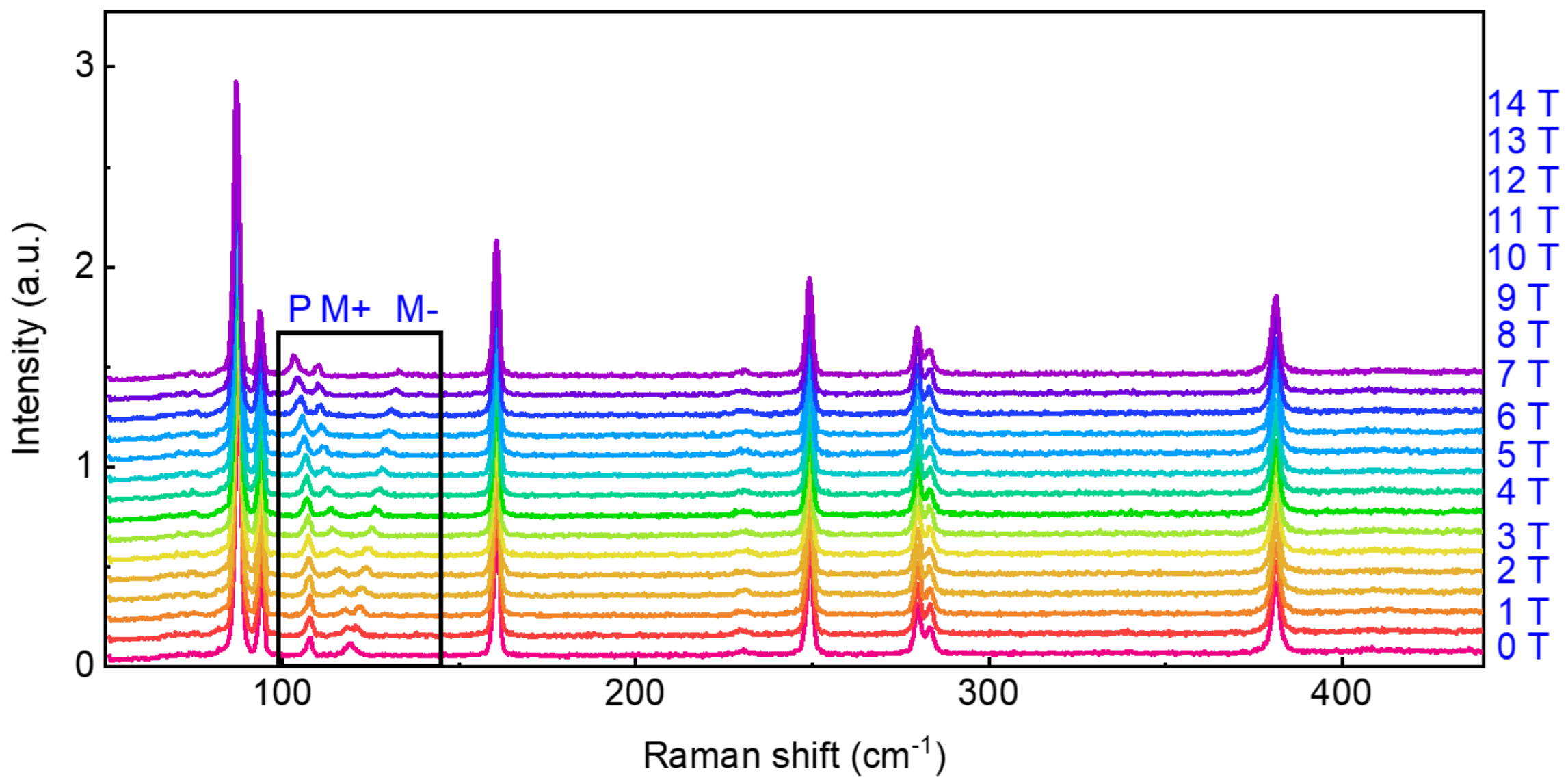

**Fig. S11.** Raw Raman spectra at 70 K under various magnetic fields. The black box indicates the energy range of interest. The P, M+, and M- peaks represent the phonon and split magnon modes under an applied magnetic field.

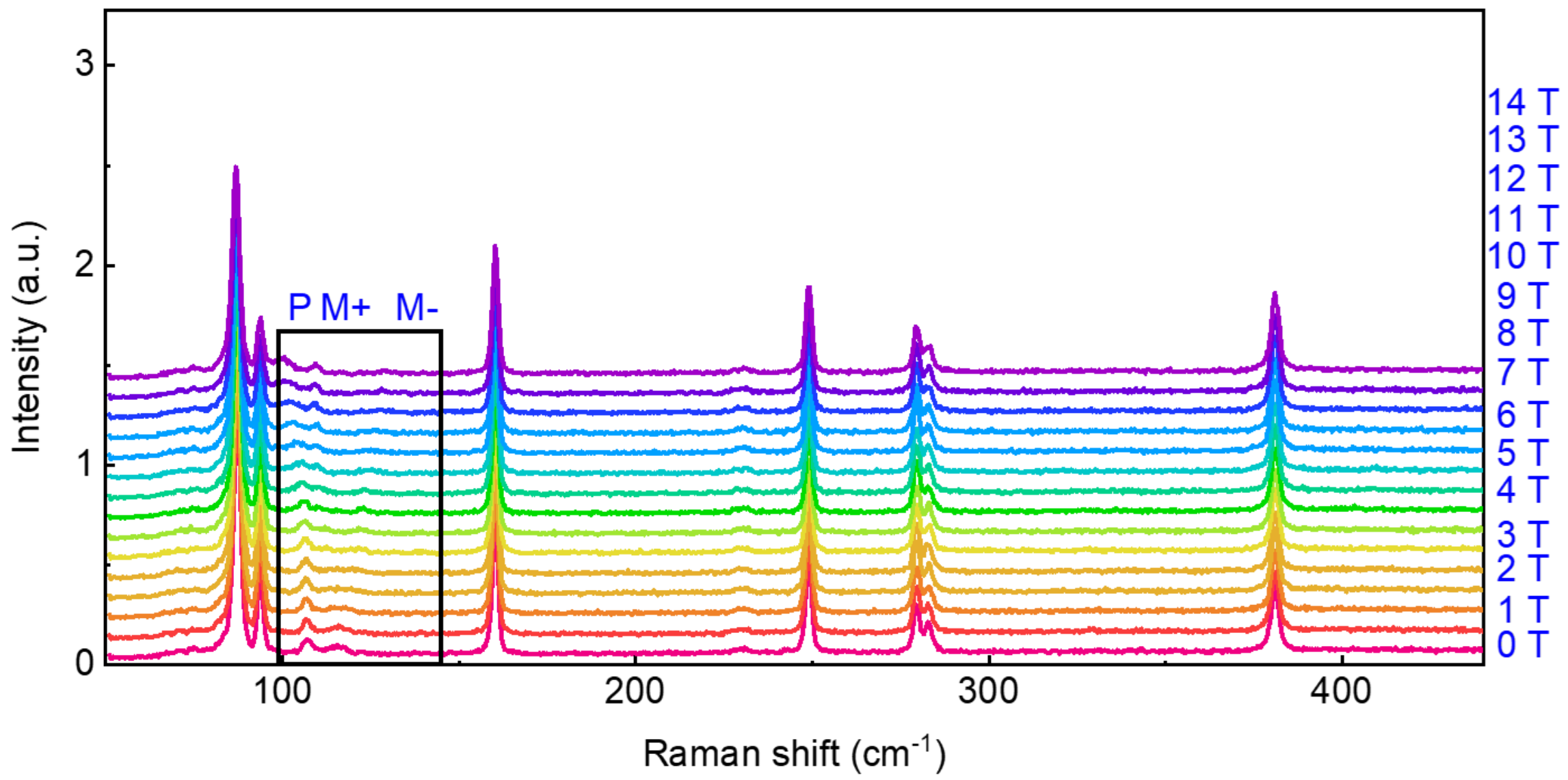

**Fig. S12.** Raw Raman spectra at 80 K under various magnetic fields. The black box indicates the energy range of interest. The P, M+, and M- peaks represent the phonon and split magnon modes under an applied magnetic field.

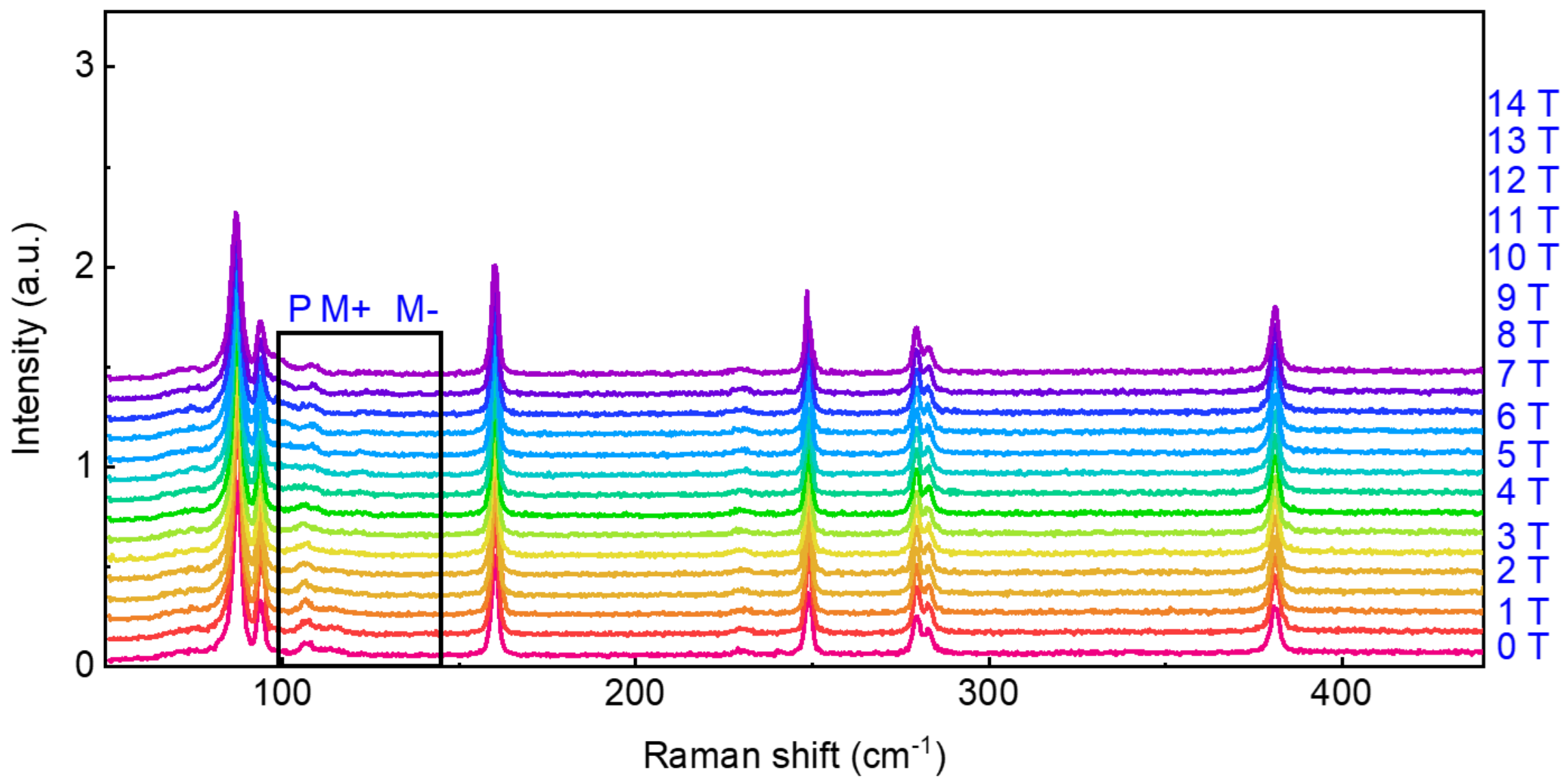

**Fig. S13.** Raw Raman spectra at 90 K under various magnetic fields. The black box indicates the energy range of interest. The P, M+, and M- peaks represent the phonon and split magnon modes under an applied magnetic field.

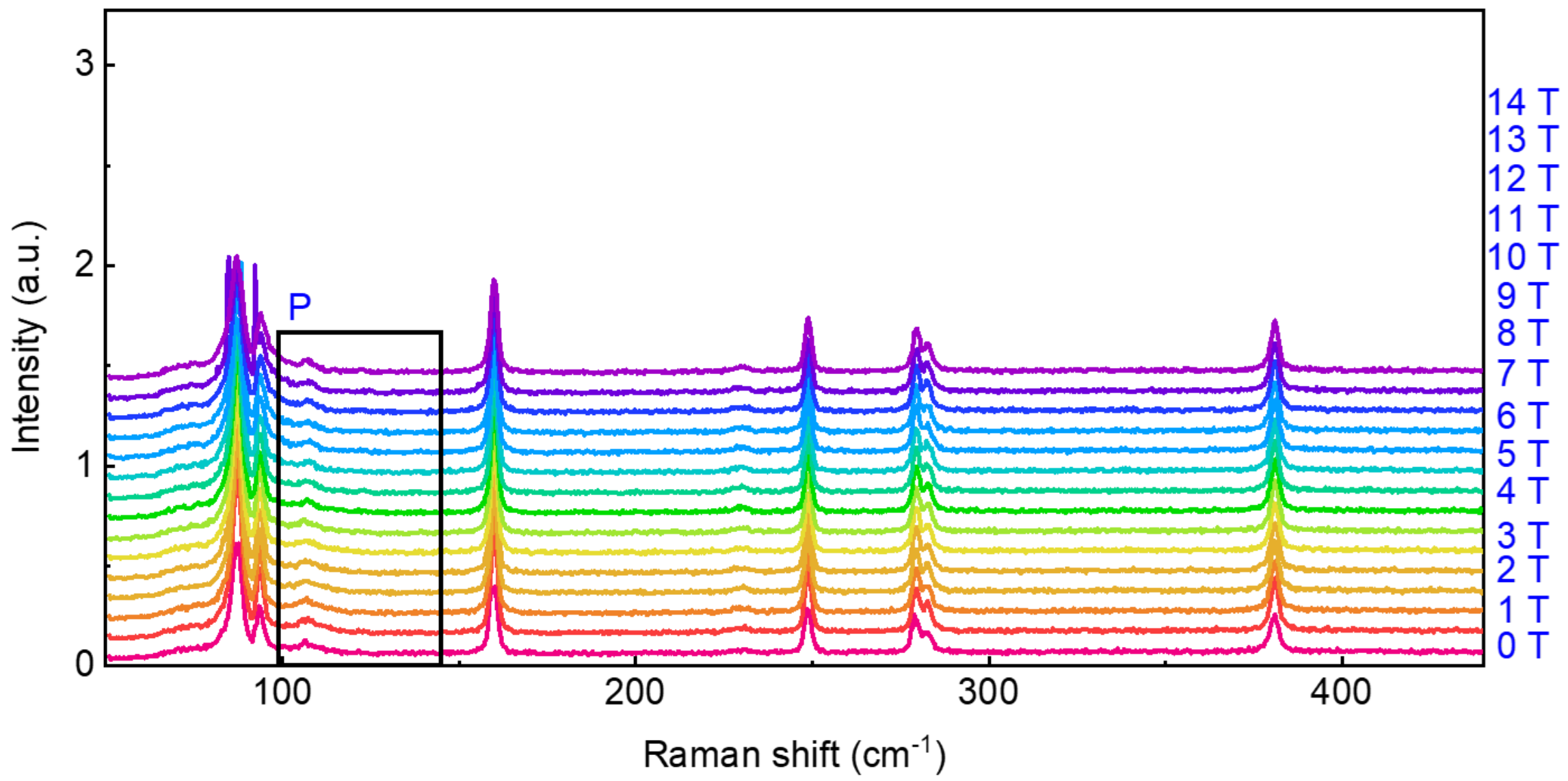

**Fig. S14.** Raw Raman spectra at 100 K under various magnetic fields. The black box indicates the energy range of interest. The P peak represents the phonon mode, and no magnon peak is observed.

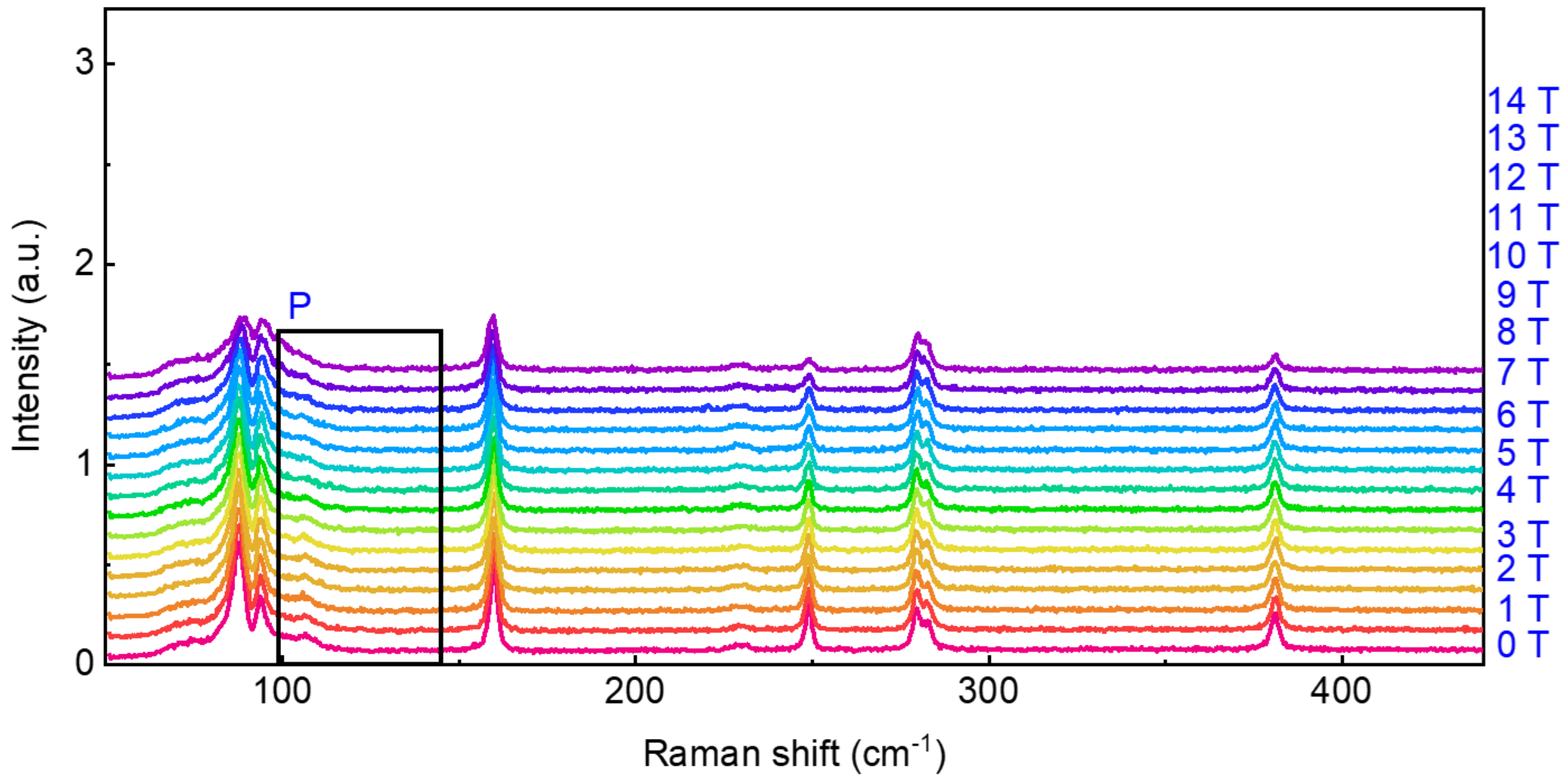

**Fig. S15.** Raw Raman spectra at 110 K under various magnetic fields. The black box indicates the energy range of interest. The P peak represents the phonon mode, and no magnon peak is observed.

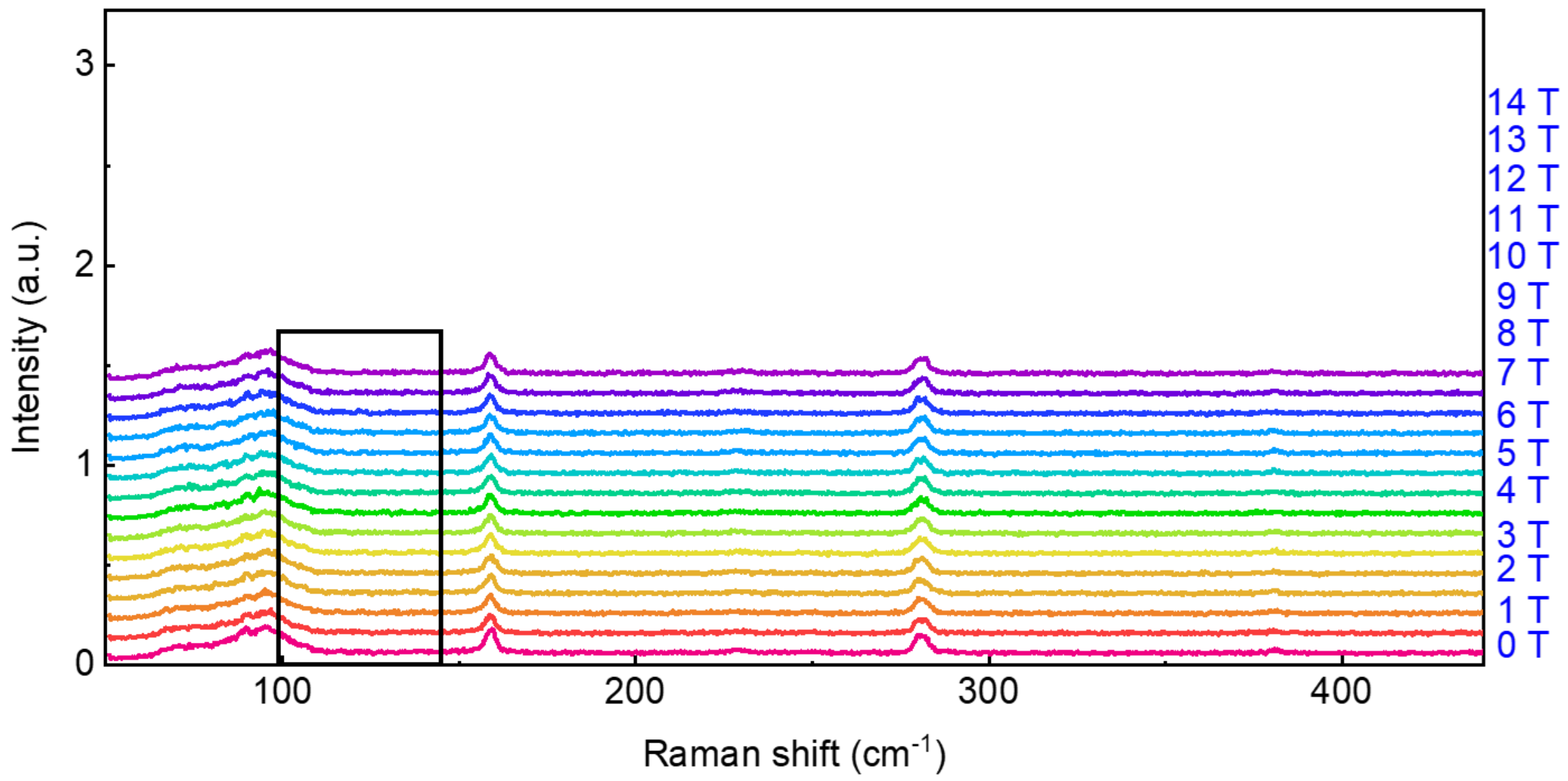

**Fig. S16.** Raw Raman spectra at 120 K under various magnetic fields. The black box indicates the energy range of interest, within which no peak is observed.

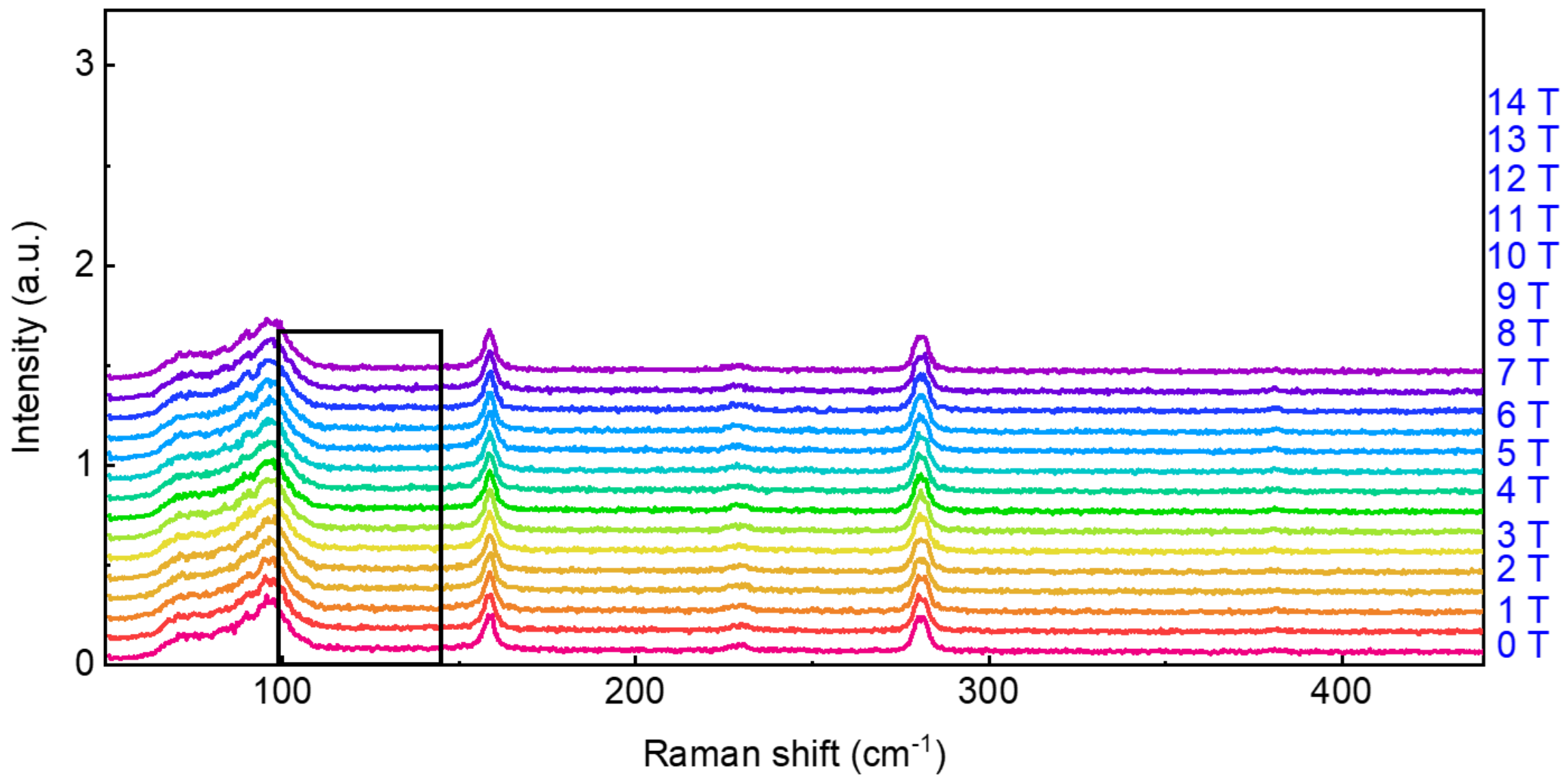

**Fig. S17.** Raw Raman spectra at 130 K under various magnetic fields. The black box indicates the energy range of interest, within which no peak is observed.